\documentclass[a4paper,10pt]{article}

\usepackage[left=2.4cm,right=2.4cm,top=2.8cm,bottom=2.8cm]{geometry}

\usepackage[utf8]{inputenc}
\usepackage[T1]{fontenc}
\usepackage{amsfonts,amssymb,amsthm,mathtools}
\usepackage{siunitx}
\usepackage{soul,xcolor}
\usepackage[round]{natbib}
\usepackage{subcaption}

\definecolor{dB}{rgb}{.0,.5,.4}

\newcommand\pt{\partial}%.....partial derivative
\newcommand\grad{\vec{\nabla}}%.....gradient
\renewcommand\i{i}%.....imaginary unit
\renewcommand\d{\text{d}}%.....differential
\renewcommand\Re{\operatorname{Re}}%.....real part
\renewcommand\Im{\operatorname{Im}}%.....imaginary part

\newcommand\bfactor{b}%.....corrective factor

\begin{document}

\title{Two hydrodynamic effects allow strongly nonlinear cochlear response with level-independent admittance}

\author{Renata Sisto$^1$ \and Daniele Belardinelli$^{1,}$\thanks{belardinellid@gmail.com} \and Arturo Moleti$^2$}
\date{$^1$INAIL, Department of Medicine, Epidemiology and Hygiene, Monte Porzio Catone (RM), Italy. \\ $^2$Physics Department, University of Rome Tor Vergata, Rome, Italy.}

%\author{Renata Sisto}
%\author{Daniele Belardinelli}
%\email{belardinellid@gmail.com}
%\affiliation{INAIL, Department of Medicine, Epidemiology and Hygiene, Monte Porzio Catone (RM), Italy.}
%
%\author{Arturo Moleti}
%\affiliation{Physics Department, University of Rome Tor Vergata, Rome, Italy.}

\maketitle

\renewcommand{\abstractname}{}
\begin{abstract}
This paper discusses the role of 2-D/3-D cochlear fluid hydrodynamics in the generation of the large nonlinear dynamical range of the basilar membrane (BM) and pressure response, in the decoupling between cochlear gain and tuning, and in the dynamic stabilization of the high-gain BM response in the peak region. The large and closely correlated dependence on stimulus level of the BM velocity and fluid pressure gain \citep{dong2013detection}, is consistent with a physiologically-oriented schematization of the outer hair cell (OHC) mechanism if two hydrodynamic effects are accounted for: amplification of the differential pressure associated with a focusing phenomenon, and viscous damping at the BM-fluid interface. The predictions of the analytical 2-D WKB approach are compared to solutions of a 3-D finite element model, showing that these hydrodynamic phenomena yield stable high-gain response in the peak region and a smooth transition among models with different effectiveness of the active mechanism, mimicking the cochlear nonlinear response over a wide stimulus level range. This study explains how an effectively anti-damping nonlinear OHC force may yield large BM and pressure dynamical ranges along with an almost level-independent admittance.
\end{abstract}

\section{Introduction}

The extraordinary dynamic range and sharp tuning of the human peripheral auditory system has challenged modeling attempts both before and after the discovery of the underlying physiological mechanisms. Although several characteristics of the cochlear active filter have been successfully modeled in the last decades, a few aspects of the experimentally observed phenomenology remain puzzling.

The experimental basilar membrane (BM) response is a peaked nonlinear function close to the characteristic frequency (CF) place, and linear elsewhere. Here, linear means that the response grows proportionally to the stimulus level, and, normalizing the response to the stimulus level, one gets a set of “sensitivity” or “gain” curves, of decreasing peak level (and increasing bandwidth) with increasing stimulus level. We will define the nonlinear dynamical range of the response as the peak gain difference between the lowest and the highest experimental stimulus levels (in some cases, exceeding \SI{50}{\decibel}, \citealp{rhode2007basilar}). For practical reasons, such experiments are typically performed at a specific place along the BM varying the stimulus frequency. The cochlear scaling symmetry suggests that the same behavior is expected by measuring the spatial dependence of the BM response at a fixed frequency. In the experiments, a shift of the peak of the BM response to lower frequencies is observed with increasing stimulus level, which would correspond to a basal shift of the peak of the spatial profile.

Measurements by \citet{dong2013detection} showed similar nonlinear dynamical ranges for the BM velocity (about \SI{45}{\decibel}) and the differential pressure (about \SI{35}{\decibel}), with comparatively little nonlinear variation of their ratio (\SI{10}{\decibel}). In other words, the large nonlinear variation of the BM peak gain is not associated with a corresponding variation of the local admittance, defined here as the ratio between the basilar membrane velocity and the local pressure.

Although the nonlinear dynamical range of the BM gain is qualitatively accompanied by a significant variation of the BM response bandwidth and phase (slope), this variation is partially decoupled from the nonlinear gain behavior, i.e., the nonlinear change of the gain is not accompanied by a correspondingly large variation of the bandwidth (tuning), or of the otoacoustic emission delay. This may be partly due to the local and instantaneous nonlinear cochlear response, which is not equivalent to that of a set of linear systems of different gain and tuning, set by the value of the stimulus level \citep{sisto2015decoupling}. Nevertheless, this behavior may also be considered as another evidence that the nonlinear variation of the admittance of the system in the resonant region cannot be the unique responsible for the whole gain dynamical range, because in that case a tighter coupling between gain and tuning could have been predicted.

The peaked and nonlinear BM response within a strongly dissipative system, such as the cochlea, is generally explained by the presence of an active amplification loop localized in the outer hair cells (OHC) system, capable of injecting mechanical power in the vibrating system, counteracting the viscous losses. The main cochlear nonlinearity is localized in the opening probability of the OHC MET channels and the related nonlinear stiffness of the hair bundle \citep{avan2013auditory}, so it should be possible to trace back any nonlinear behavior of the BM response to the OHC nonlinear response. As the OHC current is activated by the tilt of the stereocilia, it is natural to assume that the OHC force is functionally dependent on the reticular lamina (RL) displacement relative to the tectorial membrane (TM). More generally, a closed loop system in which the additional OHC force is generated by, and amplifies the motion of different elements of the Organ of Corti (OC) is a natural assumption of several cochlear models.

A class of linear 1-D transmission-line models was developed (e.g., \citealp{zweig1991finding,talmadge1998modeling}), in which the differential pressure between the scalae acts as driving force on the BM transverse vibration, which is coupled to the fluid velocity and pressure fields through incompressibility. This coupling yields the cochlear slow traveling waves, in which transverse BM vibration and differential pressure propagate forward and backward along the longitudinal direction. In this view, viscous losses are represented by a damping term on the BM proportional to its velocity, and the OHC active forces are represented by explicit or effective (i.e., in phase with velocity) anti-damping terms. The OHC nonlinearity was introduced in 1-D transmission line models by assuming instantaneous nonlinear dependence of the additional OHC active force on the local BM displacement/velocity level, requiring therefore a solution in the time domain (e.g., \citealp{moleti2009otoacoustic,sisto2010different}). The introduction of a second mechanical degree of freedom, associated to either the TM or the RL, led to the development of two-degrees of freedom (2DOF) linear and nonlinear models (e.g., \citealp{neely1986model,sisto2019constraints}), which allow one to predict the motion of different parts of the OC, still within the limits of the 1-D formulation.

In the 1-D transmission line formulation, the additional OHC active force directly modifies the local response to the driving pressure by modifying the transverse local admittance of the BM, introducing effective anti-damping terms in the local oscillator equation. Such 1-D models proved capable of explaining the experimental high-gain, sharply tuned response of the BM, yielding, in some cases, a region of overall negative damping, at the rather high price of assuming a fine-tuned balance between maximal active anti-damping terms and passive damping terms, to prevent instability.

Cochlear models including a physiologically-oriented schematization of the OHCs (e.g., \citealp{lu2006fast}) yield a nonlinear force proportional to the displacement of specific elements of the OC, which, passing through the low-pass filter of the slow membrane voltage build-up, becomes an effective anti-damping force in the basal cochlea \citep{sisto2021low}. This interesting behavior is related to the spatial dependence of the cell conductance and capacitance \citep{nam2012optimal}, which implies that the local low-pass cutoff frequency is increasingly lower than the local characteristic frequency (CF) approaching the cochlear base, and becomes comparable to it or higher only near the apex. A simple 2DOF 1-D cochlear model \citep{sisto2021low}, including a plausible schematization of the OHC piezoelectric actuation and low-pass filtering, shows that, in the mid-to-high frequency range, the competition between increasing low-pass attenuation and increasing low-pass phase-shift yields relatively small gain dynamics (\SIrange{15}{20}{\decibel}) and sharp tuning above \SI{1}{\kilo\hertz} and almost no gain dynamics and poor tuning below \SI{1}{\kilo\hertz}, recalling some features of the so-called apical-basal transition. Trying to increase the gain of such a model by strengthening the OHC effective anti-damping effect leads into unstable parameter regions, thus a different additional amplification mechanism seems necessary to explain both the non-null (although small, about \SIrange{15}{20}{\decibel}) nonlinear dynamical range of the apical cochlea and the much larger one of the basal cochlea.

In this paper, we show that by expanding 1-D cochlear models to include at least one additional spatial dimension (perpendicular to the BM surface), one may account for two hydrodynamic effects that help explaining the observed phenomenology:

1) The 2-D hydrodynamical phenomenon of “pressure focusing” contributes significantly to the gain dynamical range in the peak region \citep{shera2005coherent}, without changing significantly the admittance \citep{altoe2020cochlear}, i.e., the ratio between BM velocity and pressure. Although focusing is a linear mechanism, it depends on the amplitude of the wavenumber in the peak region, which is a function of the effectiveness of the nonlinear active mechanism, hence, of the stimulus level. This focusing effect explains a consistent fraction of the total nonlinear gain dynamical range. Consequently, a moderate nonlinear variation of the local admittance is expected in the presence of a much larger BM gain dynamical range.

2) Fluid viscosity in a “focused” pressure and velocity field provides a correction to the admittance of a particular functional form, which allows stable solutions with high nonlinear gain of the BM response, and further decreases the nonlinear variation of the admittance in an increasingly active system. Indeed, we will show that, as the gain and the wavenumber in the peak region increase with decreasing stimulus level, the viscous damping correction to the admittance increases proportionally to the wave number, partially balancing the growth of the increasingly effective OHC anti-damping term.

In the WKB framework, this phenomenology may be interpreted as follows: as each frequency component of the traveling wave (TW) approaches its resonant cochlear region, the real part of the wavenumber $k$ (proportional to the reciprocal of the wavelength) increases dramatically. In this short-wave region ($\Re(k)H>1$, with $H$ being the typical scalae height) the wavenumber becomes proportional to the admittance (e.g., \citealp{siebert1974ranke,zweig2015linear}), whereas in the long-wave region it is proportional to its square root. In the short-wave regime, the pressure is focused, due to the volume conservation, in a thin layer close to the BM, within a distance of the order of the TW wavelength. The pressure focusing phenomenon boosts both the BM response and the pressure in the peak region, without changing their local ratio (i.e., the local admittance). This hydrodynamic focusing phenomenon appears responsible for a considerable part of the total gain of the TW near the tonotopic place, and of its dependence on level, frequency, and location \citep{altoe2020cochlear}. The higher gain at low stimulus levels would still be a (partly indirect) consequence of the OHC active mechanism, because in a low-$Q$ passive cochlea the short-wave condition would be weakened, and the focusing mechanism would be less effective, and because an effective active mechanism compensates for the large viscous damping forces, which, as we will show analytically, increase with increasing wavenumber. In any case, the simultaneous boost of both components of the TW means that one does not need a large tuning/admittance nonlinear variation to explain the measured large nonlinear dependence on stimulus level of the BM velocity gain.

In this study, we analyze the two above-mentioned physical mechanisms, finding analytical expressions valid in the WKB approximation. We then discuss the results of numerical simulations. A very simple 3-D finite element model of the cochlea, in which the Navier-Stokes equations are solved for a viscous incompressible fluid coupled to the mechanical equations for an orthotropic elastic membrane, was designed to capture the essence of the effect of the focusing and viscous dissipation phenomena on the TW propagation and amplification. A 2-D transmission line model was also designed to implement the analytical expressions found in the WKB approximation, to demonstrate that the phenomenology of the 3-D model is sufficiently well captured by the analytical expressions.

\section{Analytic treatment of pressure focusing and viscosity in the 2-D cochlear model}

Let us recall the basic equations of the traveling-wave propagation along the cochlea in a simplified 2-D box model of length $L$ and (semi-)height $H$, starting from the linear Navier-Stokes equations written for an incompressible viscous fluid with mass density $\rho$ and bulk viscosity $\mu$:
\begin{equation}%l
\grad\cdot\vec{u} = 0, \label{eq:incompressibility}
\end{equation}%l
\begin{equation}%l
\rho\frac{\pt\vec{u}}{\pt t} = - \grad p + \mu\nabla^2\vec{u}, \label{eq:linearNS}%l
\end{equation}
where $\vec{u}$ is the fluid velocity and $p$ is the pressure. The Laplacian viscous contribution in the linear Navier-Stokes equation~(\ref{eq:linearNS}) comes from the divergence of the viscous stress tensor $\mu[\grad\vec{u}+(\grad\vec{u})^\intercal]$ (the superscript $^\intercal$ means transposition), in which the incompressibility condition~(\ref{eq:incompressibility}) has been used (e.g.,
\citealp{landau1987fluid,batchelor2000introduction}).

Let $x$ and $z$ be the cochlear longitudinal axis coordinate and the coordinate orthogonal to the (undeformed) BM, respectively, and let us indicate with $u$ and $w$ respectively the $x$ and $z$ components of the fluid velocity: $\vec{u}\equiv(u,w)$. The impenetrability conditions on the box rigid walls read
\begin{equation*}
u(z=\pm H) = 0.
\end{equation*}
The boundary on the BM assumes a deformable shape localized at $z=\xi(x,t)$. The impenetrability condition at the BM then requires:
\begin{equation*}
w(z=\xi) = u(z=\xi)\frac{\pt\xi}{\pt x} + \dot{\xi},
\end{equation*}
where $\dot{\xi}\equiv\pt\xi/\pt t$. The fluid incompressibility implies:
\begin{equation*}
\frac{\pt u}{\pt x} + \frac{\pt w}{\pt z} = 0.
\end{equation*}
Integrating along $z$ one gets:
\begin{equation*}
\int_{\xi}^H \frac{\pt u}{\pt x} \d z = - \int_{\xi}^H \frac{\pt w}{\pt z} \d z = w(z=H) - w(z=\xi) = u(z=\xi)\frac{\pt\xi}{\pt x} + \dot{\xi}.
\end{equation*}
But
\begin{equation*}
\int_{\xi}^H \frac{\pt u}{\pt x} \d z = \frac{\pt}{\pt x}\int_{\xi}^H u \d z + u(z=\xi)\frac{\pt\xi}{\pt x},
\end{equation*}
thus
\begin{equation}%l
\frac{\pt}{\pt x}\int_{\xi}^H u \d z = \dot{\xi}. \label{eq:volumeconservation}
\end{equation}%l
Equation~(\ref{eq:volumeconservation}) is exact, in other words, the volume conservation is independent of the ratio between $\xi$ and $H$, which is obviously very small.

From Eq.~(\ref{eq:volumeconservation}) the propagation equation for the pressure averaged over the $z$ axis is straightforwardly obtained. Neglecting that the integration starts at $\xi$, that is using that $|\xi|\ll H$, and deriving with respect to time:
\begin{equation}%l
\frac{\pt}{\pt x}\int_0^H \frac{\pt u}{\pt t} \d z = \ddot{\xi}, \label{eq:conservation}
\end{equation}%l
and, neglecting bulk viscous terms in Eq.~(\ref{eq:linearNS}):
\begin{equation}%l
\frac{\pt^2}{\pt x^2}\int_0^H p_{\text{d}} \d z = - 2\rho\ddot{\xi};\label{eq:keyequation}
\end{equation}%l
the factor 2 is due to considering the differential pressure $p_{\text{d}} = p(z)-p(-z)$ between the scalae, which is the only part of the pressure field involved in the slow traveling wave. From now on, we will always refer to this odd part of the pressure field. Equation~(\ref{eq:keyequation}) relates the BM transverse acceleration to the second spatial derivative of the differential pressure with respect to the longitudinal direction integrated along the vertical axis. We can then write
\begin{equation}%l
\frac{\pt^2\overline{p_{\text{d}}}}{\pt x^2} = - \frac{2\rho}{H}\ddot{\xi}, \label{eq:keyequationaverage}
\end{equation}%l
where $\overline{p_{\text{d}}}(x,t)$ is the differential pressure averaged along the vertical axis.

In a 2-D cochlear box model, the differential pressure satisfying the equations and the boundary conditions for an incompressible fluid can be approximated in the frequency domain by basis functions of the form \citep{shera2005coherent,duifhuis2012cochlear}:
\begin{equation}%l
p_{\text{d}}(x,z,\omega) \propto f(x,\omega)\frac{\cosh(k(H-z))}{\cosh(kH)} e^{-\i\int_0^x k(x',\omega) \d x'}. \label{eq:pressureWKB}
\end{equation}%l
Here, we use the approximation proposed by \citet{shera2005coherent}:
\begin{equation*}
f(x,\omega) = \sqrt{\frac{k(0,\omega)}{k(x,\omega)}}.
\end{equation*}

In the WKB approximation, by integrating Eq.~(\ref{eq:pressureWKB}), the pressure averaged along the vertical axis is related to the local pressure near the BM by:
\begin{equation}%l
\overline{p_{\text{d}}} = \frac{p_{\text{d}}(z=0)}{\alpha}, \label{eq:focusing}
\end{equation}%l
with
\begin{equation}%l
\alpha = \frac{kH}{\tanh(kH)}. \label{eq:focusingfactor}
\end{equation}%l

Note that, although the focusing factor and the wave vector are both complex, the BM and pressure response drop by several orders of magnitude in the cochlear region where the imaginary part of the wave vector becomes of the same order as the real part. In the same region, the WKB approximation also breaks down because the spatial derivative of the real part of the wave number becomes negative (see, e.g., \citealp{talmadge1998modeling}). Therefore, for practical purposes, the amplitude of the wave vector and its real part may be approximated by each other.

If we neglect the effect of bulk viscosity in Eq.~(\ref{eq:linearNS}), the even part of the $z$ component of the fluid velocity can be derived from Eq.~(\ref{eq:pressureWKB}):
\begin{equation*}
w(x,z,\omega) + w(x,-z,\omega) = - \frac{1}{\i\omega\rho}\frac{\pt p_{\text{d}}}{\pt z} = \frac{k}{\i\omega\rho}\frac{\sinh(k(H-z))}{\cosh(kH)}p_{\text{d}}(x,0,\omega). \label{eq:zvelocityWKB}
\end{equation*}
We will make use of this expression in computing the viscous contribution to the stress (force per unit surface) on the BM.

Let us recall here the second cochlear basic equation, i.e., the time-domain differential equation describing the dynamics of the single nonlinear oscillator driven by the fluid differential pressure:
\begin{equation}%l
\sigma_{\text{bm}}(\ddot{\xi}+\Gamma(x,\xi,\dot{\xi})\dot{\xi}+\omega_{\text{bm}}^2(x,\xi,\dot{\xi})\xi) = - p_{\text{d}}(x,0,t), \label{eq:BMequation0}
\end{equation}%l
where $\sigma_{\text{bm}}$ is the BM surface density, and the first member could be replaced, in a more general case, by a differential operator applied to $\xi$, or, equivalently, to $\dot{\xi}$. The oscillator in Eq.~(\ref{eq:BMequation0}) is parametric; that is, the damping coefficient and resonance frequency may be nonlinear functions of the BM displacement and velocity. In particular, the function $\Gamma(x,\xi,\dot{\xi})$ may consist of a linear passive damping term, $\Gamma_{\text{p}}(x)$, assumed to be a scale-invariant function of $x$ only, and of an active nonlinear term, $\Gamma_{\text{a}}(x,\xi,\dot{\xi})$ which schematizes the anti-damping effect of the OHC forces. In the numerical simulations of this study, this nonlinear dependence will be neglected, for simplicity, considering a set of linear models with different effectiveness of the active mechanism, roughly representing the response of the system at different stimulus levels.

From Eq.~(\ref{eq:BMequation0}), linearized and written in the frequency domain, one can define the BM admittance as the ratio between the BM velocity and the local focused differential pressure:
\begin{equation*}
\dot{\xi}(x,\omega) = Y_{\text{bm}}(x,\omega)p_{\text{d}}(x,0,\omega).
\end{equation*}
The local admittance is dependent only on the locally resonant properties of Eq.~(\ref{eq:BMequation0}), and to the local displacement and velocity through the parameters $\Gamma$ and $\sigma_{\text{bm}}$. The BM acceleration may be expressed as:
\begin{equation*}
\ddot{\xi} = \i\omega Y_{\text{bm}}p_{\text{d}}.
\end{equation*}
From Eq.~(\ref{eq:keyequationaverage}), it can be demonstrated (e.g., \citealp{shera2005coherent}) that the propagation equation for the $z$-averaged differential pressure $\overline{p_{\text{d}}}(x,\omega)$ is therefore:
\begin{equation}%l
\frac{\pt^2\overline{p_{\text{d}}}}{\pt x^2} = - \frac{2\i\omega\rho Y_{\text{bm}}}{H}p_{\text{d}}(z=0) = - \frac{2\i\omega\rho Y_{\text{bm}}}{H}\alpha\overline{p_{\text{d}}} \equiv - k^2\overline{p_{\text{d}}}, \label{eq:waveEqpressure}
\end{equation}%l
where a new relation between the wavenumber and the local admittance, taking into account the 2-D fluid hydrodynamic focusing, is now given by:
\begin{equation}%l
k^2 = \frac{2\alpha\i\omega\rho}{H}Y_{\text{bm}} = \alpha k_{\text{lw}}^2. \label{eq:newwavenumber}
\end{equation}%l
Indeed, in the long-wave limit ($\Re(k)H\ll1$, $\alpha\to1$)\footnote{Formally, the only condition $\Re(k)H\ll1$ would give $\alpha\to\alpha_{\text{lw}}=\Im(k)H/\tan(\Im(k)H)$. However, in the region interested by the long-wave approximation, we have also $\Im(k)H\ll1$, implying $\alpha_{\text{lw}}\simeq1$.} the wavenumber tends to its long-wave limit:
\begin{equation}%l
k \to k_{\text{lw}} = \sqrt{\frac{2\i\omega\rho}{H}Y_{\text{bm}}}, \label{eq:lwavenumber}
\end{equation}%l
while, in the short-wave limit ($\Re(k)H\gg1$, $\alpha\to kH$):
\begin{equation*}
k \to k_{\text{sw}} = 2\i\omega\rho Y_{\text{bm}}.
\end{equation*}

\subsection{Viscosity correction to the admittance}

Given a unit vector $\vec{n}$, the viscous force exerted by the fluid on a unit surface orthogonal to $\vec{n}$, on the side where $\vec{n}$ points, is \citep{landau1987fluid,batchelor2000introduction}
\begin{equation*}
\vec{f}_{\vec{n}} = \mu[\grad\vec{u}+(\grad\vec{u})^\intercal]\cdot\vec{n}.
\end{equation*}
The $z$ component of the force exerted on a horizontal unit surface from above (below) is then obtained by choosing $\vec{n}=\hat{z}$ ($\vec{n}=-\hat{z}$):
\begin{equation*}
\hat{z}\cdot\vec{f}_{\pm\hat{z}} = \pm \mu\hat{z}\cdot\left(\grad w+\frac{\pt\vec{u}}{\pt z}\right) = \pm 2\mu\frac{\pt w}{\pt z}.
\end{equation*}
A viscous dissipation stress $S$ then acts on the BM from each side, being given by
\begin{equation}%l
S = 2\mu\left.\frac{\pt w}{\pt z}\right]_{z=0+} - 2\mu\left.\frac{\pt w}{\pt z}\right]_{z=0-} = - \frac{2\alpha\mu}{\i\omega\rho}k^2\overline{p_{\text{d}}}, \label{eq:stress}
\end{equation}%l
where Eqs.~(\ref{eq:zvelocityWKB}), and (\ref{eq:focusing}) have been used. The last passage of Eq.~(\ref{eq:stress}) neglects, for simplicity, the contribution to the derivative of $w$ associated with the rotational part of the velocity field. A rough numerical estimate, based on the 3-D WKB solutions given by \citet{steele1999cochlear} for the vector and scalar potentials of the fluid dynamic field, suggests that using the correct expression, including the contribution from the vector potential, $S$ is \numrange{2}{3} times larger, with the same dependence on the wavenumber. Therefore we introduce here a factor $b=\num{2.5}$, to account for that. Thus, using Eqs.~(\ref{eq:waveEqpressure}), and (\ref{eq:keyequationaverage}), the additional force per unit surface on the BM can be written as:
\begin{equation}%l
S = - \frac{4\alpha\bfactor\mu}{\i\omega H}\ddot{\xi} = - \frac{4\alpha\bfactor\mu}{H}\dot{\xi}. \label{eq:stresscorrected}
\end{equation}%l

Let us now write Eq.~(\ref{eq:BMequation0}) in the frequency domain adding the dissipative stress $S$ to the pressure contribution:
\begin{equation}%l
\sigma_{\text{bm}}(-\omega^2+\i\omega\Gamma+\omega_{\text{bm}}^2)\xi = - p_{\text{d}}(z=0) - \frac{4\alpha\bfactor\mu}{H}\i\omega\xi, \label{eq:BMequationcorrected}
\end{equation}%l
which can be written as
\begin{equation}%l
\sigma_{\text{bm}}\left[\omega_{\text{bm}}^2-\omega^2+\i\omega\left(\Gamma + \frac{4\alpha\bfactor\mu}{\sigma_{\text{bm}}H}\right)\right]\xi = - p_{\text{d}}(z=0), \label{eq:BMequation}
\end{equation}%l
and the corrected cochlear admittance may be defined as
\begin{equation*}
\tilde{Y}_{\text{bm}} = \frac{\i\omega\xi}{p_{\text{d}}(z=0)} = - \frac{\i\omega}{\sigma_{\text{bm}}\tilde{\Delta}},
\end{equation*}
where
\begin{equation*}
\tilde{\Delta}(x,\omega) = \omega_{\text{bm}}^2-\omega^2+\i\omega\left(\Gamma + \frac{4\alpha\bfactor\mu}{\sigma_{\text{bm}}H}\right) = \omega_{\text{bm}}^2-\omega^2+\i\omega\left(\Gamma + \frac{4\bfactor\mu k}{\sigma_{\text{bm}}\tanh(kH)}\right).
\end{equation*}

Taking into account in Eq.~(\ref{eq:newwavenumber}) the correction to the admittance due to viscosity, we find that the wavenumber must obey the following relation:
\begin{equation}%l
k^2 = \frac{2\alpha\omega^2\rho}{\sigma_{\text{bm}}H\left[\omega_{\text{bm}}^2-\omega^2+\i\omega\left(\Gamma + \frac{4\alpha\bfactor\mu}{\sigma_{\text{bm}}H}\right)\right]}. \label{eq:k2recurrence}
\end{equation}%l

In the short-wave limit this relation reduces to
\begin{equation*}
k_{\text{sw}} = \frac{2\omega^2\rho}{\sigma_{\text{bm}}\left[\omega_{\text{bm}}^2-\omega^2+\i\omega\left(\Gamma + \frac{4\bfactor\mu k_{\text{sw}}}{\sigma_{\text{bm}}}\right)\right]},
\end{equation*}
which can be solved for $k_{\text{sw}}$, while in the long-wave limit we get
\begin{equation*}
k_{\text{lw}}^2 = \frac{2\omega^2\rho}{\sigma_{\text{bm}}H\left[\omega_{\text{bm}}^2-\omega^2+\i\omega\left(\Gamma + \frac{4\bfactor\mu}{\sigma_{\text{bm}}H}\right)\right]}.
\end{equation*}

Thus, in the short-wave limit the viscous fluid provides a stabilizing damping force in the peak region, counteracting the focusing phenomenon, both effects being proportional to $\Re(k)$. This force comes from the viscous dissipation at the fluid-BM interface. This viscous effect also modifies the effective resonant frequency by an amount proportional to $\Im(k)$. Although we have assumed, for simplicity, the admittance of a harmonic oscillator transmission line, the derivation of Eq.~(\ref{eq:stresscorrected}) is independent of the choice of the particular model of the admittance one starts with. Using the simple resonant admittance form associated with Eq.~(\ref{eq:BMequation0}) allows a straightforward estimate of the size of the viscous and elastic corrections to the damping and resonant frequency.

\section{Numerical models and their solution}

\subsection{3-D Finite-Element Model}

We developed a simple finite-element (FE) model of the cochlea using Comsol Multiphysics 5.5 (COMSOL Inc., USA). Although the parameters are roughly inspired to those of the mouse cochlea, the model is not designed to reproduce any animal in particular, because the purpose of the study is to highlight, with minimal model complexity, the phenomenology of TW amplification by focusing and stabilization by viscous damping. We represented the uncoiled cochlea as a 3-D box of height $2H$, width $W$, and length $L$, partitioned at $z = 0$ by an elastic membrane (BM) of thickness $d$ and density $\rho_{\text{bm}}$ The scalae are filled by an incompressible viscous fluid of viscosity $\mu$ and density $\rho$. As before, the variable $x$ is the longitudinal axis, and $z$ is the direction orthogonal to the BM. The solid domain consists of a shell element for the BM, modeled as an orthotropic material in which the Young modulus in the direction parallel to the BM axis is negligible ($E_x \ll E_y = E_z$), and the Poisson's modulus is $\nu$. A non-structured tetrahedral mesh with element minimum size $s_{\text{min}}$ was used. The exponential longitudinal dependence of the Young modulus was set to reproduce a realistic tonotopic map $\omega_{\text{bm}}(x)$.

We introduced damping on the BM assuming a passive damping coefficient equal to the local resonance angular frequency, $\Gamma_{\text{p}}(x)=\omega_{\text{bm}}(x)$, and schematized the active OHC forces by including an additional anti-damping load on the BM:
\begin{equation*}
p_{\text{ohc}}(x) = \sigma_{\text{bm}}\Gamma_{\text{a}}(x)\dot{\xi} = \i\omega G\omega_{\text{bm}}(x)\sigma_{\text{bm}}\xi.
\end{equation*}
The coefficient of the active term is proportional to a multiplicative factor, $G$, whose value was varied in the range $\numrange{0.25}{2.75}$ to simulate the dependence on stimulus level of the effectiveness of the nonlinear OHC force, similarly to that assumed in \citet{wang2016cochlear} for the same purpose. This is a crude schematization of nonlinearity, because the actual nonlinear OHC response is dependent on the local excitation level and not on any global parameter such as the stimulus level. Whereas model responses are low-$Q$ and passive-like for $G\leq1$, the peak gain is increased by about \SI{30}{\decibel} for $G=\num{2.75}$.

Because this study focuses on the fundamental physical nature of two hydrodynamic effects, we ignored the internal details of the OC. We note, however, that the actual movements of the complex structures of the OC within a viscous fluid are likely to increase the size of the viscous losses substantially, as shown by \citet{prodanovic2019power}, who take explicitly into account viscous losses in the OC. For this reason, we also performed simulations with a larger coefficient of viscosity, ten times that of water. In that case, the coefficient $G$ of the anti-damping term was varied between 1 and \num{4.5}, to get a range of BM and pressure response similar to that obtained in the same model using the nominal viscosity of water.

The FE model is solved in the frequency domain. The acoustic field is completely solved for both pressure and fluid velocity, using the Comsol Thermoacoustics routines, which implement the complete set of Navier-Stokes equations. The BM is coupled to the fluid by means of the Thermoacoustics fluid–structure Multiphysics interaction. The complex differential pressure is calculated by measuring the pressure along the central ($y = 0$) axis parallel to the BM axis in the fluid layer immediately above and immediately below the BM (nominally $z=0\pm$, but actually $z_{\pm}=\SI{\pm15}{\micro\meter}$, due to the finite size of the mesh elements). The local admittance is calculated as the complex ratio between the BM velocity and the local differential pressure. The main parameters of the FE model are listed in Table \ref{tab:FEparameters}.

\begin{table}[ht]
\centering
\caption{\label{tab:FEparameters}Main parameters of the FE model.}
\begin{tabular}{c}
\\
\hline\hline
$H = \SI{d-3}{\meter}$\\
\hline
$L = \SI{3.2d-2}{\meter}$\\
\hline
$W = \SI{4d-4}{\meter}$\\
\hline
$d = \SI{2d-5}{\meter}$\\
\hline
$\rho_{\text{bm}} = \rho = \SI{1000}{\kilogram\per\meter\cubed}$\\
\hline
$\mu = \mu_{\text{water}},\num{10}\mu_{\text{water}}$\\
\hline
$\mu_{\text{water}} = \SI{0.7d-3}{\pascal\second} \ (\SI{37}{\celsius})$\\
\hline
$s_{\text{min}} = \SI{1.5d-5}{\meter}$\\
\hline
$E_x(x) = \SI{d4}{\pascal}$\\
\hline
$E_y(x) = E_z(x) = \num{7.5d8}e^{-\num{276}x}\si{\pascal}$\\
\hline
$\nu = \num{0.3}$\\
\hline
$\omega_{\text{bm}}(x) = \num{5d5}e^{-\num{138}x}\si{\radian\per\second}$\\
\hline
$\Gamma_{\text{p}}(x) = \omega_{\text{bm}}(x)\si{\radian\per\second}$\\
\hline
$\Gamma_{\text{a}}(x) = -G\omega_{\text{bm}}(x)\si{\radian\per\second}$\\
\hline\hline
\end{tabular}
\end{table}

\citet{shera2005coherent} provide a useful expression to compute the wavenumber as a function of the computed position along the BM:
\begin{equation}%l
k^2(x,\omega) = - \frac{\dot{\xi}(x,\omega)}{\int_x^L\d x'\int_{x'}^L\d x''\dot{\xi}(x'',\omega)}. \label{eq:k2Shera}
\end{equation}%l
We will use this model-independent expression to compare the results of the FE simulation with the analytical expressions of the wavenumber computed in the WKB approximation.

\subsection{Numerical computation of the WKB solutions}

WKB calculations were performed for the simple 2-D linear box model described by Eqs.~(\ref{eq:keyequationaverage}), (\ref{eq:focusing}), and (\ref{eq:BMequation}), in which a phenomenological anti-damping term of variable intensity, set by a multiplying factor $G$, was used to roughly simulate the behavior of the OHC additional force at different stimulus levels. The main parameters of the model, listed in Table \ref{tab:WKBparameters}, were chosen in order to get a a behavior, as much as possible, equivalent to that of the corresponding 3-D FE model. The viscosity coefficient was set also in this case to two different values, $\mu_{\text{water}}$ and $\num{10}\mu_{\text{water}}$.

\begin{table}[ht]
\centering
\caption{\label{tab:WKBparameters}Main parameters of the 2-D WKB model.}
\begin{tabular}{c}
\\
\hline\hline
$H = \SI{d-3}{\meter}$\\
\hline
$L = \SI{3.2d-2}{\meter}$\\
\hline
$\sigma_{\text{bm}} = \SI{0.06}{\kilogram\per\meter\squared}$\\
\hline
$\rho = \SI{1000}{\kilogram\per\meter\cubed}$\\
\hline
$\mu = \mu_{\text{water}},\num{10}\mu_{\text{water}}$\\
\hline
$\mu_{\text{water}} = \SI{0.7d-3}{\pascal\second} \ (\SI{37}{\celsius})$\\
\hline
$\omega_{\text{bm}}(x) = \num{4.2d5}e^{-\num{138}x}\si{\radian\per\second}$\\
\hline
$\Gamma_{\text{p}}(x) = \omega_{\text{bm}}(x)\si{\radian\per\second}$\\
\hline
$\Gamma_{\text{a}}(x) = -G\omega_{\text{bm}}(x)\si{\radian\per\second}$\\
\hline\hline
\end{tabular}
\end{table}

In the WKB approximation, the analytical solution is derived in the frequency domain from the assumed form of the wavenumber as a function of frequency and position. An iterative procedure, which rapidly converges, yields the value of the wavenumber consistent with Eq.~(\ref{eq:k2recurrence}). One starts by assuming that $\alpha = 1$ and $\mu = 0$, computes $k = k_{\text{lw}}$ from Eq.~(\ref{eq:lwavenumber}), then estimates $\alpha$ from Eq.~(\ref{eq:focusingfactor}) and uses the result in Eq.~(\ref{eq:newwavenumber}) to obtain a new estimate of $k^2$, and hence of $\alpha$ again from Eq.~(\ref{eq:focusingfactor}). Then one further corrects the admittance adding the viscous term and computes $k^2$ using Eq.~(\ref{eq:k2recurrence}). The new value of $k$ yields a new $\alpha$ and a new viscosity correction to the admittance. This procedure is repeated until convergence is reached (the value of $\alpha$ at the BM response peak changes by less than a specified value, e.g., \SI{1}{\decibel}).

\section{Results}

\begin{figure}[ht]
\centering
\begin{subfigure}{.32\textwidth}
\includegraphics[width=\linewidth]{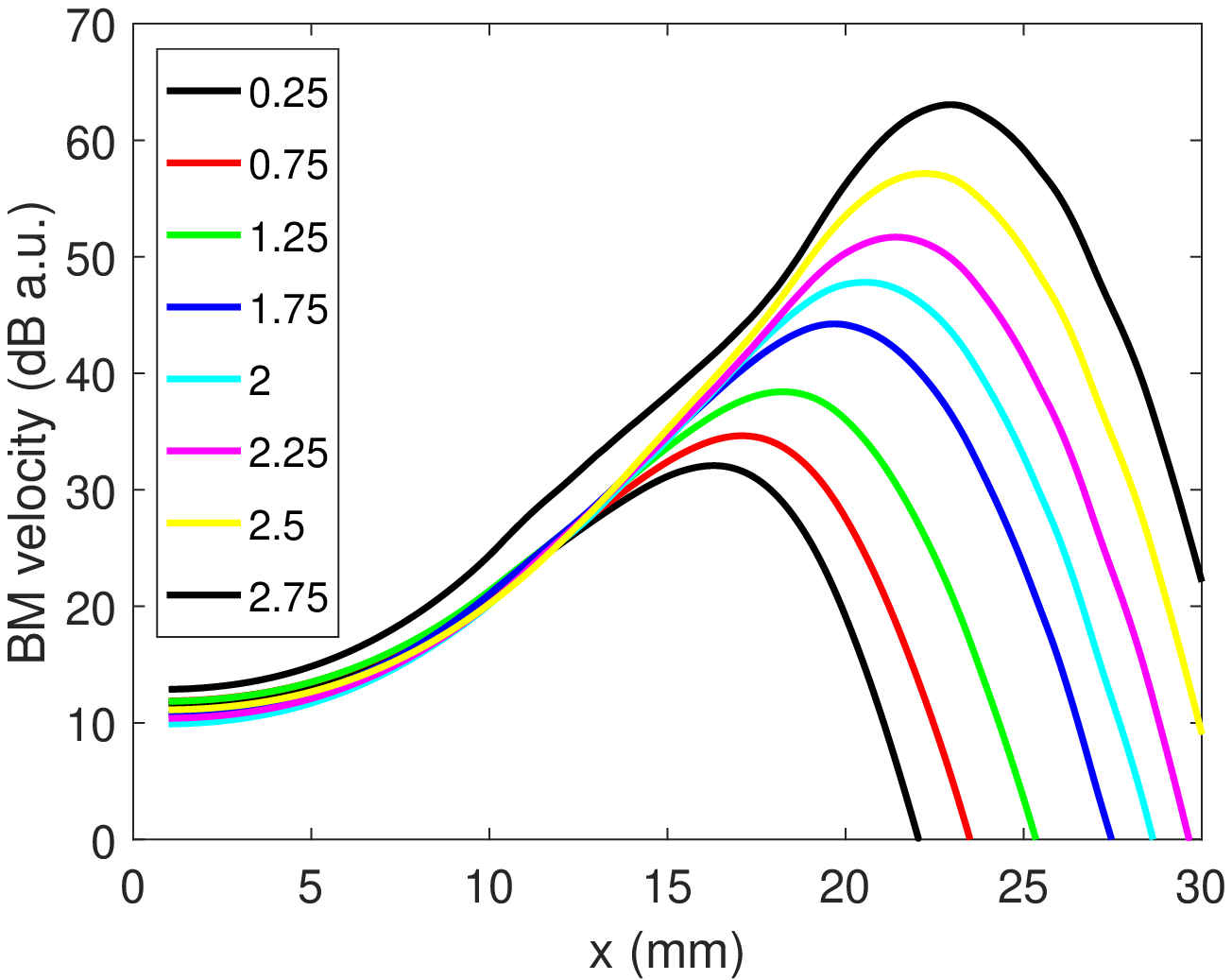}\caption{\label{fig:1a}}
\end{subfigure}
\begin{subfigure}{.32\textwidth}
\includegraphics[width=\linewidth]{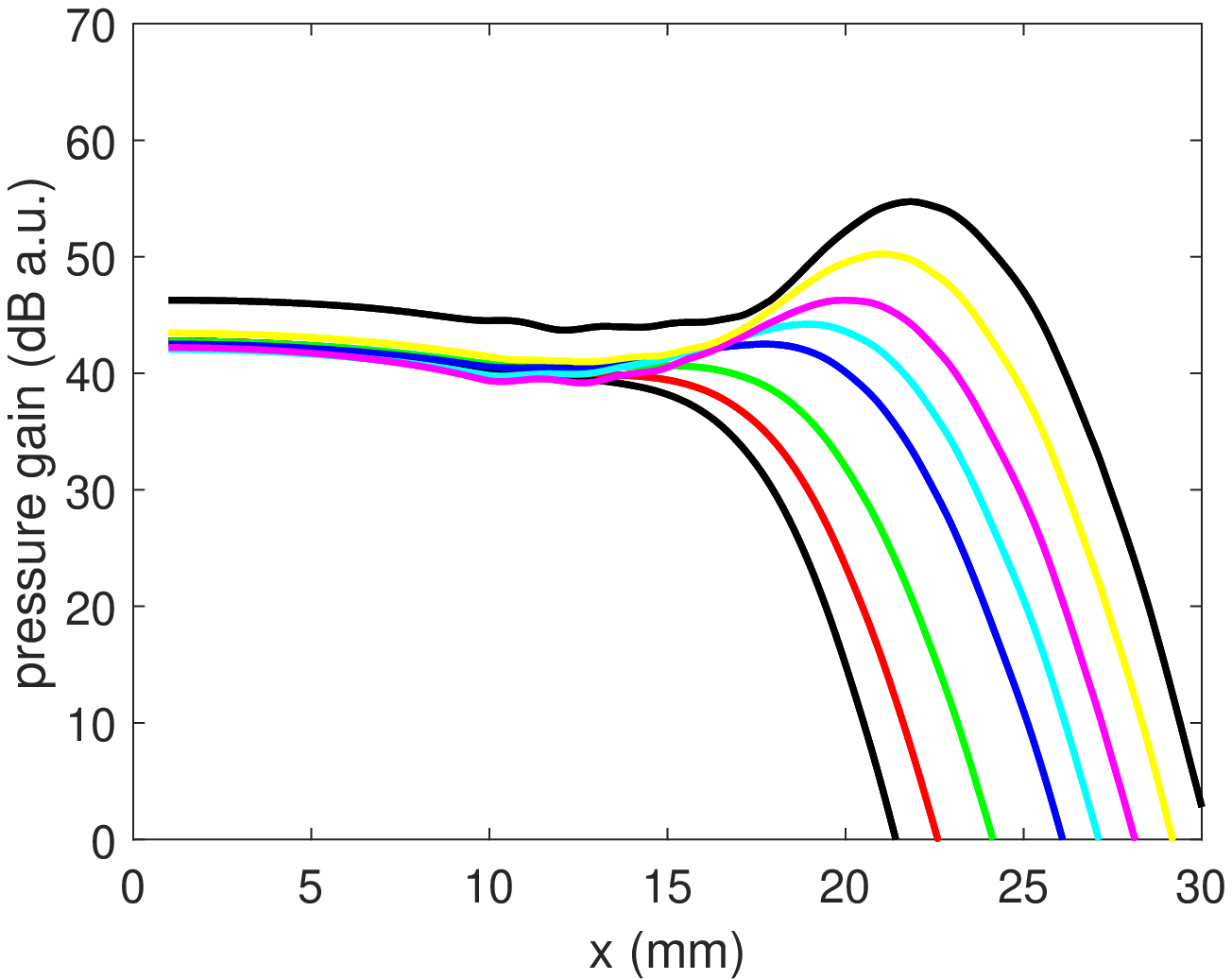}\caption{\label{fig:1b}}
\end{subfigure}
\begin{subfigure}{.32\textwidth}
\includegraphics[width=\linewidth]{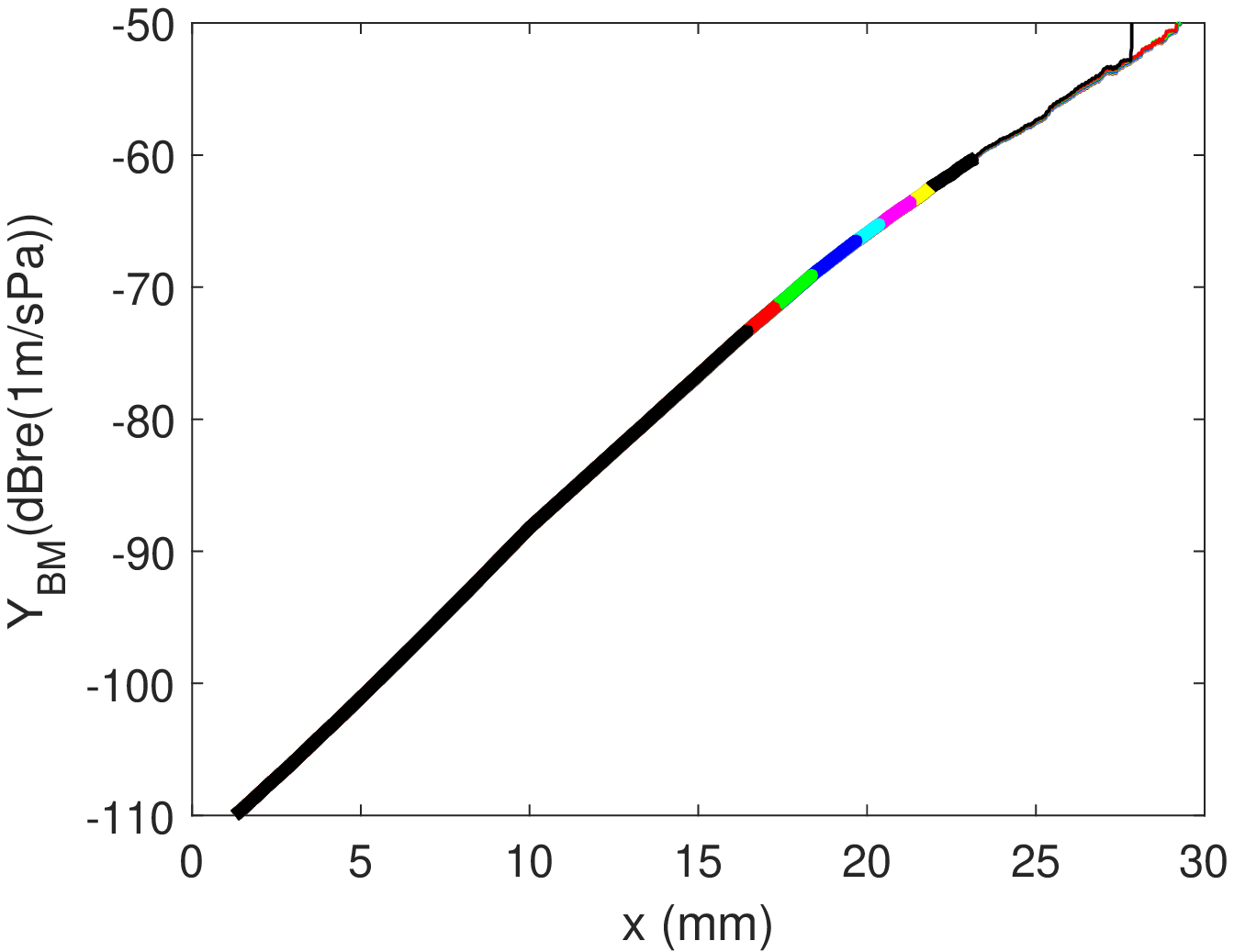}\caption{\label{fig:1c}}
\end{subfigure}
\begin{subfigure}{.32\textwidth}
\includegraphics[width=\linewidth]{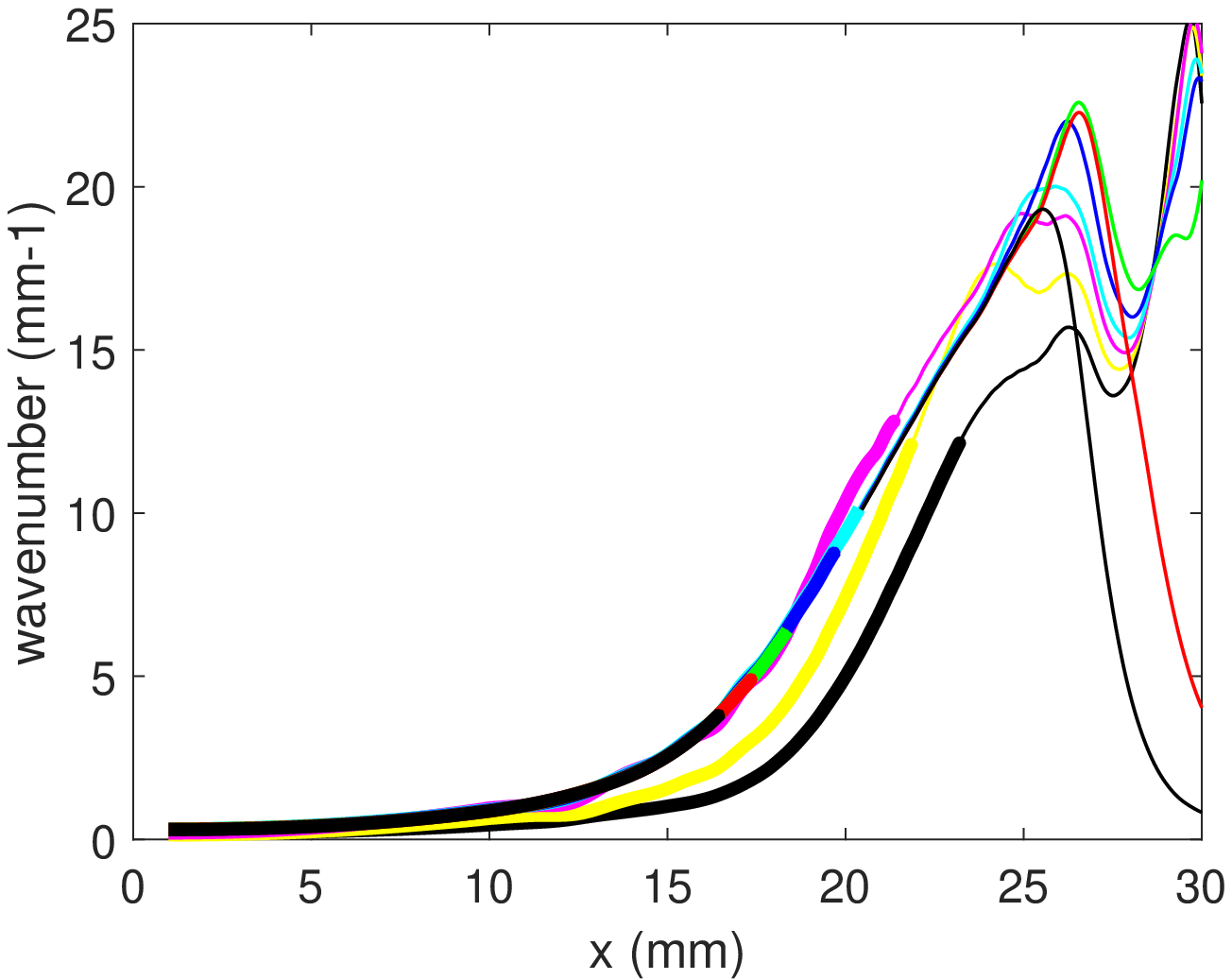}\caption{\label{fig:1d}}
\end{subfigure}
\begin{subfigure}{.32\textwidth}
\includegraphics[width=\linewidth]{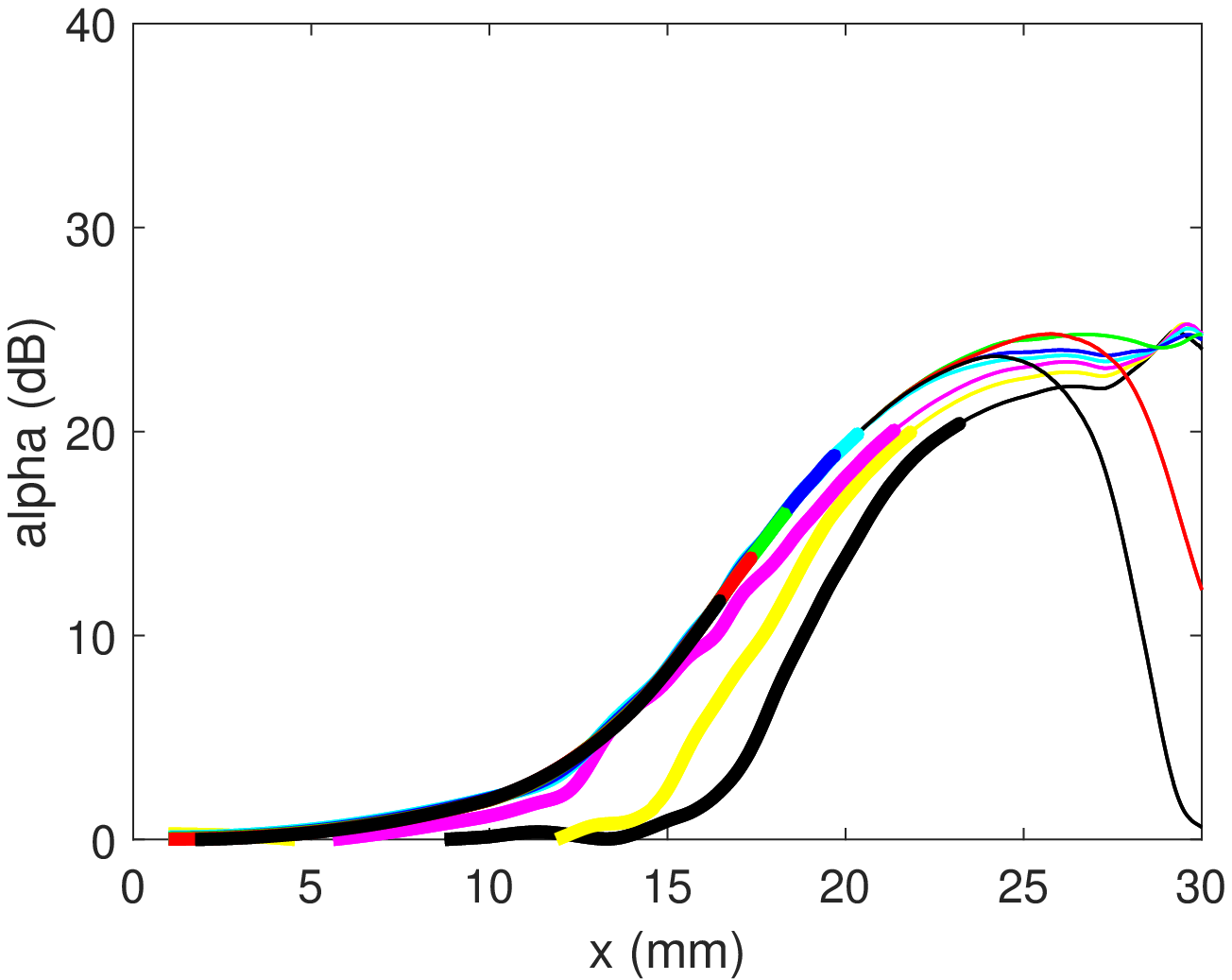}\caption{\label{fig:1e}}
\end{subfigure}
\begin{subfigure}{.32\textwidth}
\includegraphics[width=\linewidth]{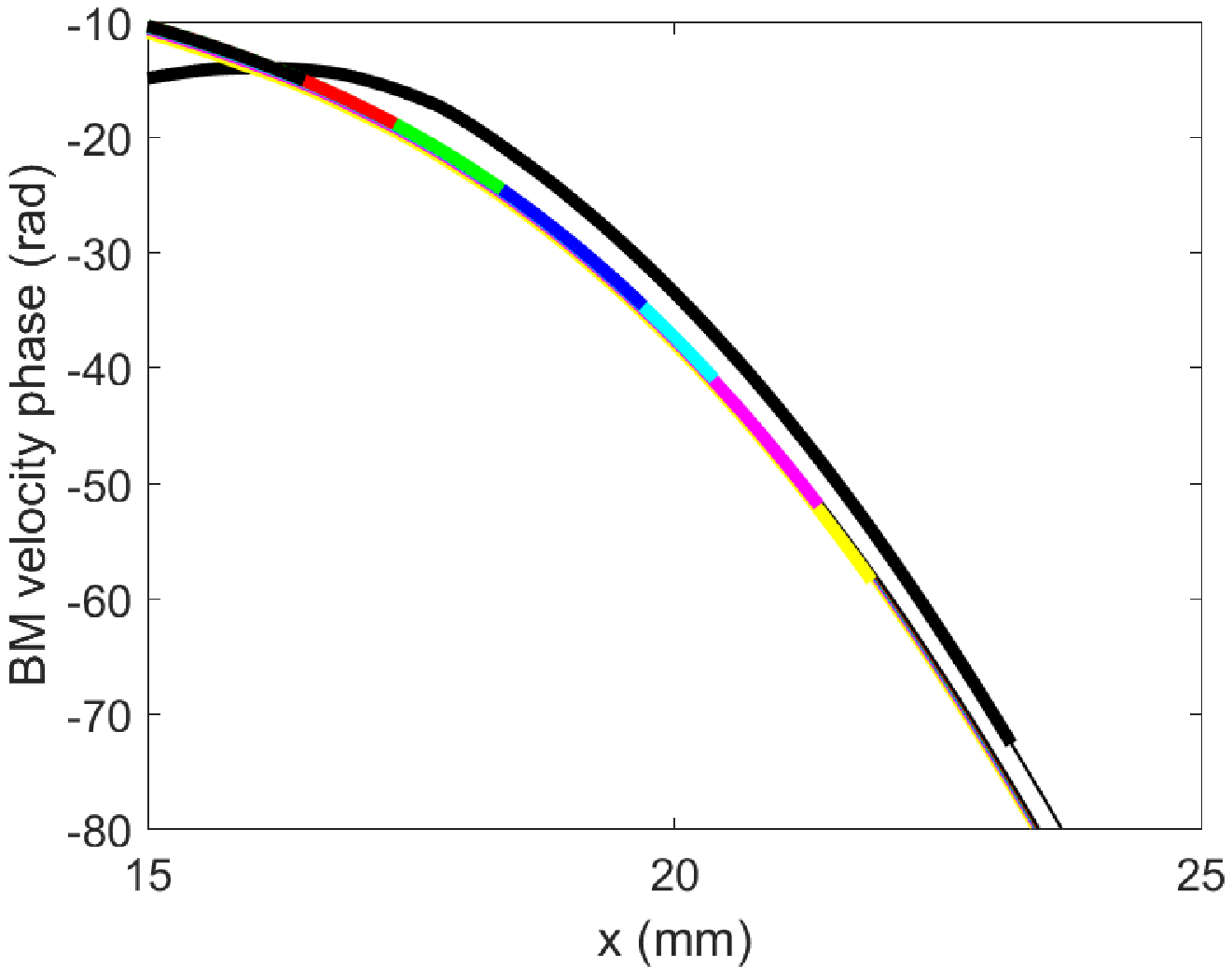}\caption{\label{fig:1f}}
\end{subfigure}
\caption{\label{fig:1}Response of the FE model for a sinusoidal stimulus of frequency \SI{2200}{\hertz}, with $\mu=\mu_{\text{water}}$; the active term factor $G$ is varied from \num{0.25} to \num{2.75}: (a) BM velocity gain profile; (b) spatial profile of the differential pressure evaluated near the BM ($z_{\pm}=\SI{\pm15}{\micro\meter}$); (c) local admittance; (d) local wavenumber computed from the BM response using the wavenumber obtained from Eq.~(\ref{eq:k2Shera}); (e) local amplitude of the factor $\alpha$; (f) BM phase. Thick lines in panels (c)-(f) extend from the base to the position of the BM response peak.}
\end{figure}
Figure~\ref{fig:1} shows the response of the FE model, with $\mu = \mu_{\text{water}}$, and the active term proportional to the multiplicative factor $G$, which is varied from $\num{0.25}$ to $\num{2.75}$. The BM gain (Fig.~\ref{fig:1a}) increases in the peak region as the active term increases, and the position of the peak shifts progressively towards the apex. The differential pressure (Fig.~\ref{fig:1b}) shows peak gain dynamics smaller than that of the BM velocity, despite the invariance of the admittance profile, because, with increasing $G$, the peak moves to regions of higher admittance. This is shown in Fig.~\ref{fig:1c}, where the admittance is plotted using a thin line beyond the position of the response peak. The same representation is used to show the spatial dependence of the wavenumber amplitude (Fig.~\ref{fig:1d}), of the focusing factor alpha Fig.~\ref{fig:1e}, and of the BM phase Fig.~\ref{fig:1f}. Although the local values of all these physical quantities are almost insensitive of the model activity factor $G$, the admittance and the factor $\alpha$ at the peak of the response increases by more than \SI{10}{\decibel} with increasing $G$. Therefore, the focusing factor and the admittance at the peak of the response are both dependent on the strength of the active nonlinear mechanism.

The large change of the phase lag and slope in the peak region, which is not consistent with physiological group delays (both neural and otoacoustic), and the large apical shift of the response peak with increasing $G$, suggests that the oversimplified schematization of the OC may have underestimated the effect of fluid viscosity. Indeed, fluid viscous losses provide (see also \citealp{prodanovic2019power}) a sharp cutoff to the growth of the TW in the peak region, limiting both the apical shift of the response and the group delay at the peak. Therefore, we repeated our simulations using a viscosity coefficient ten times larger than that of water. This may be considered as a rather crude way to account for viscous losses within the elements of the OC \citep{prodanovic2019power}.
\begin{figure}[ht]
\centering
\begin{subfigure}{.32\textwidth}
\includegraphics[width=\linewidth]{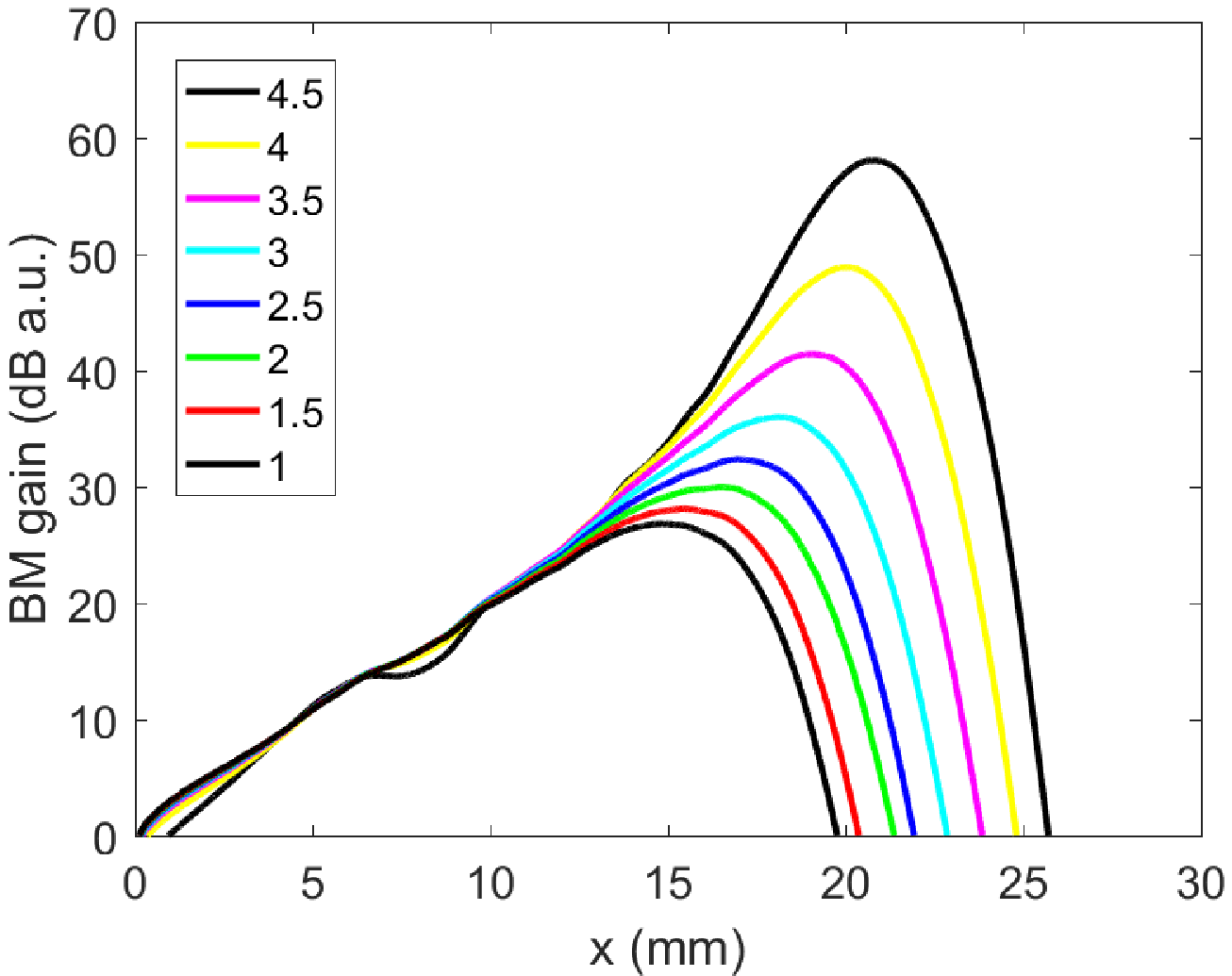}\caption{\label{fig:2a}}
\end{subfigure}
\begin{subfigure}{.32\textwidth}
\includegraphics[width=\linewidth]{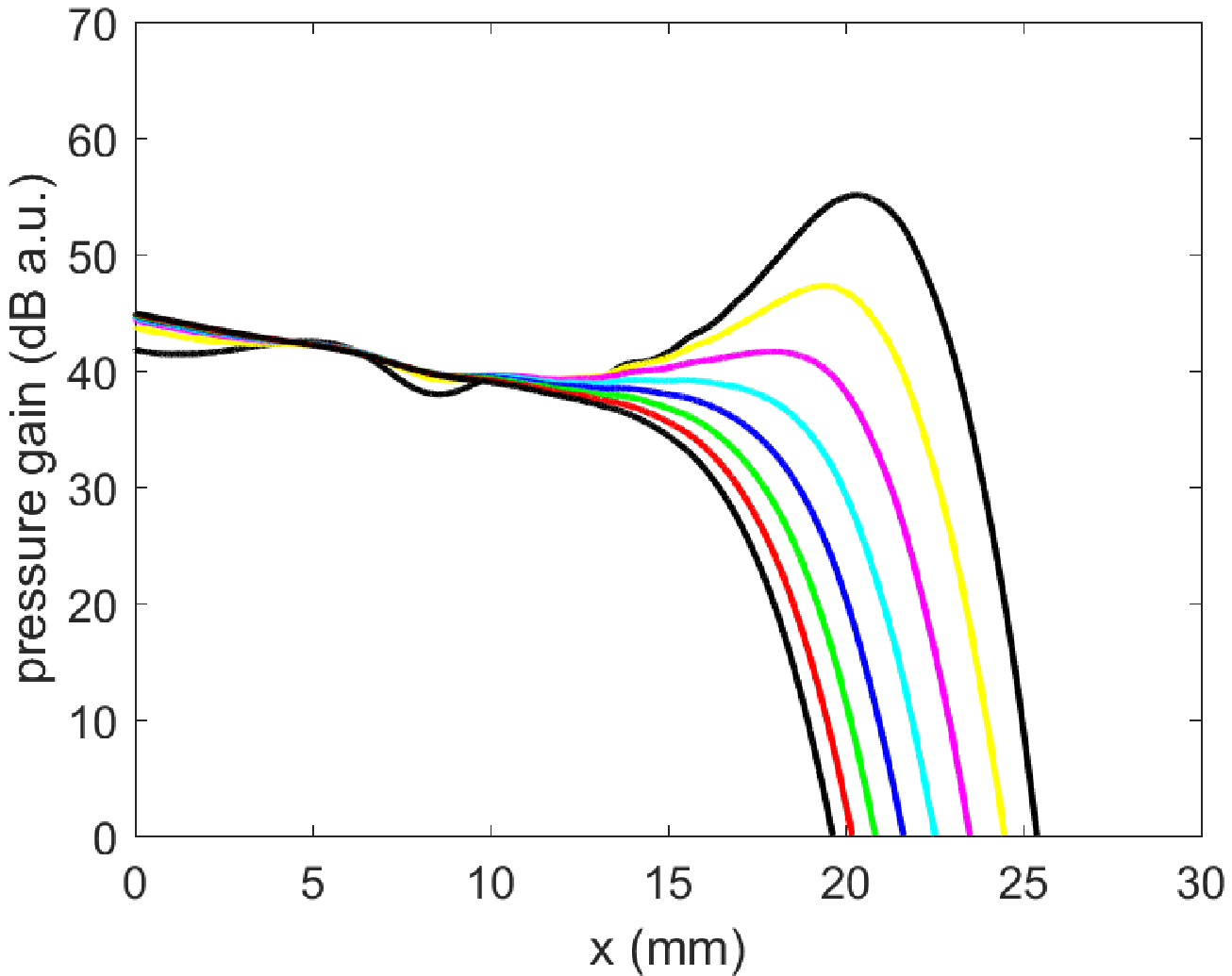}\caption{\label{fig:2b}}
\end{subfigure}
\begin{subfigure}{.32\textwidth}
\includegraphics[width=\linewidth]{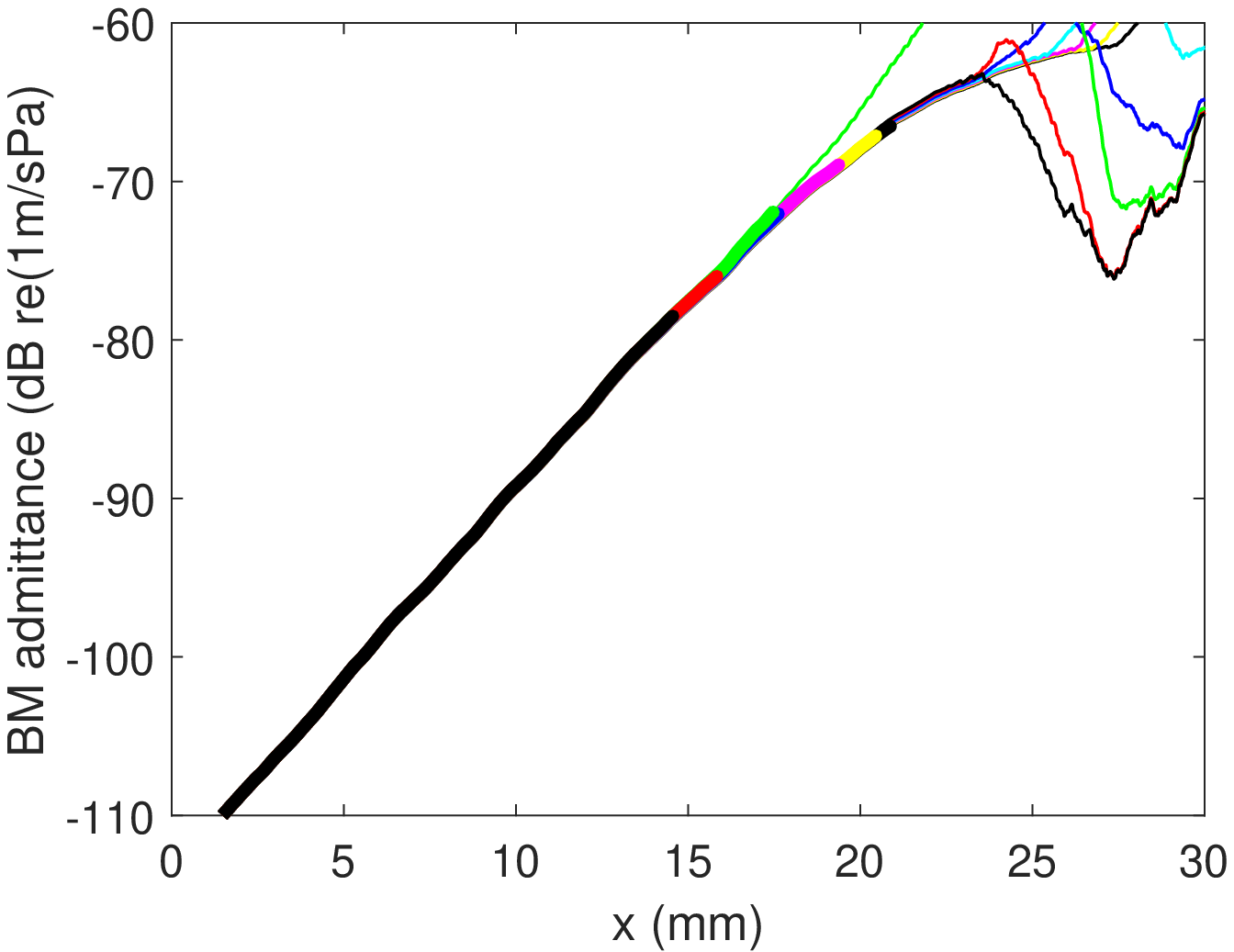}\caption{\label{fig:2c}}
\end{subfigure}
\begin{subfigure}{.32\textwidth}
\includegraphics[width=\linewidth]{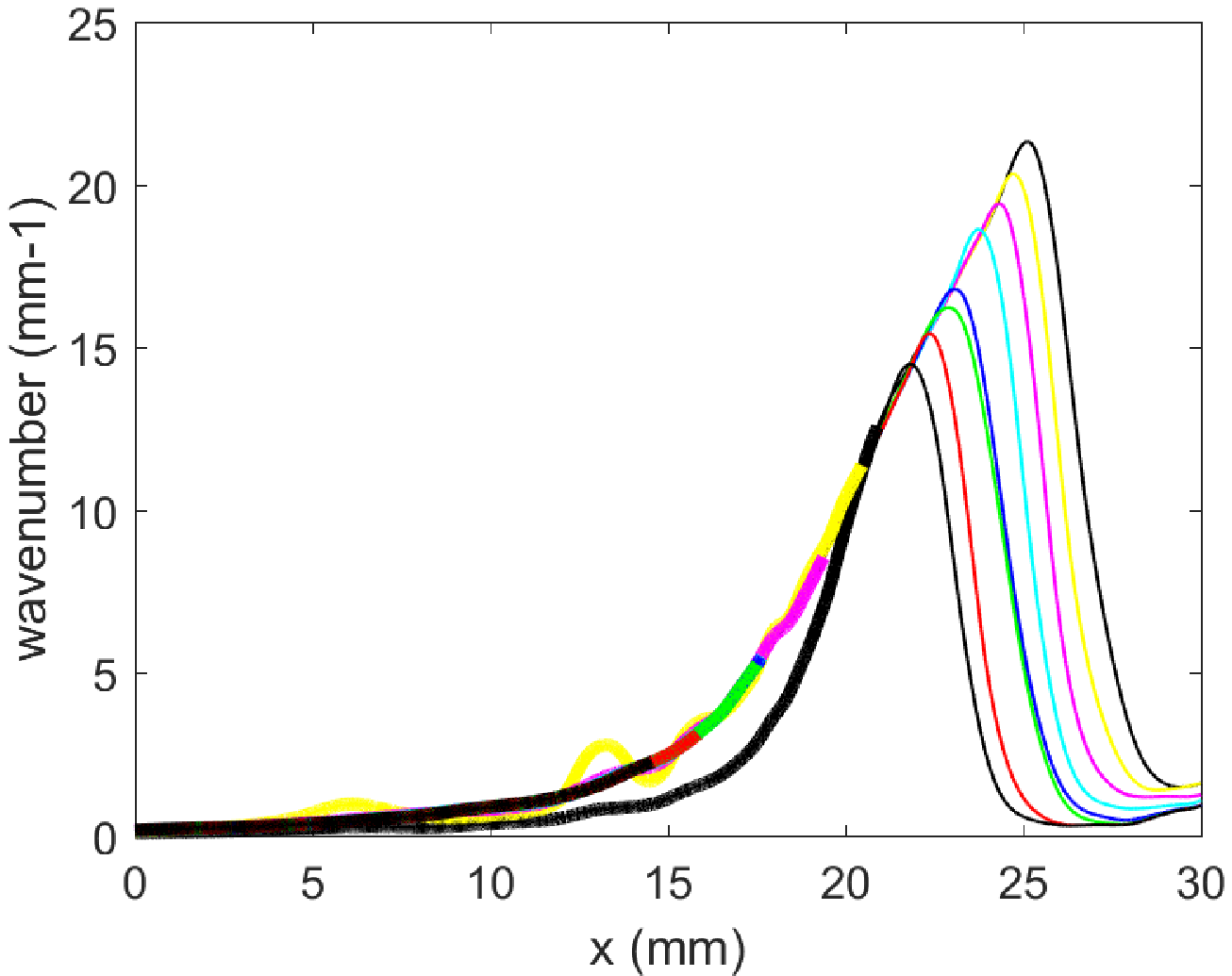}\caption{\label{fig:2d}}
\end{subfigure}
\begin{subfigure}{.32\textwidth}
\includegraphics[width=\linewidth]{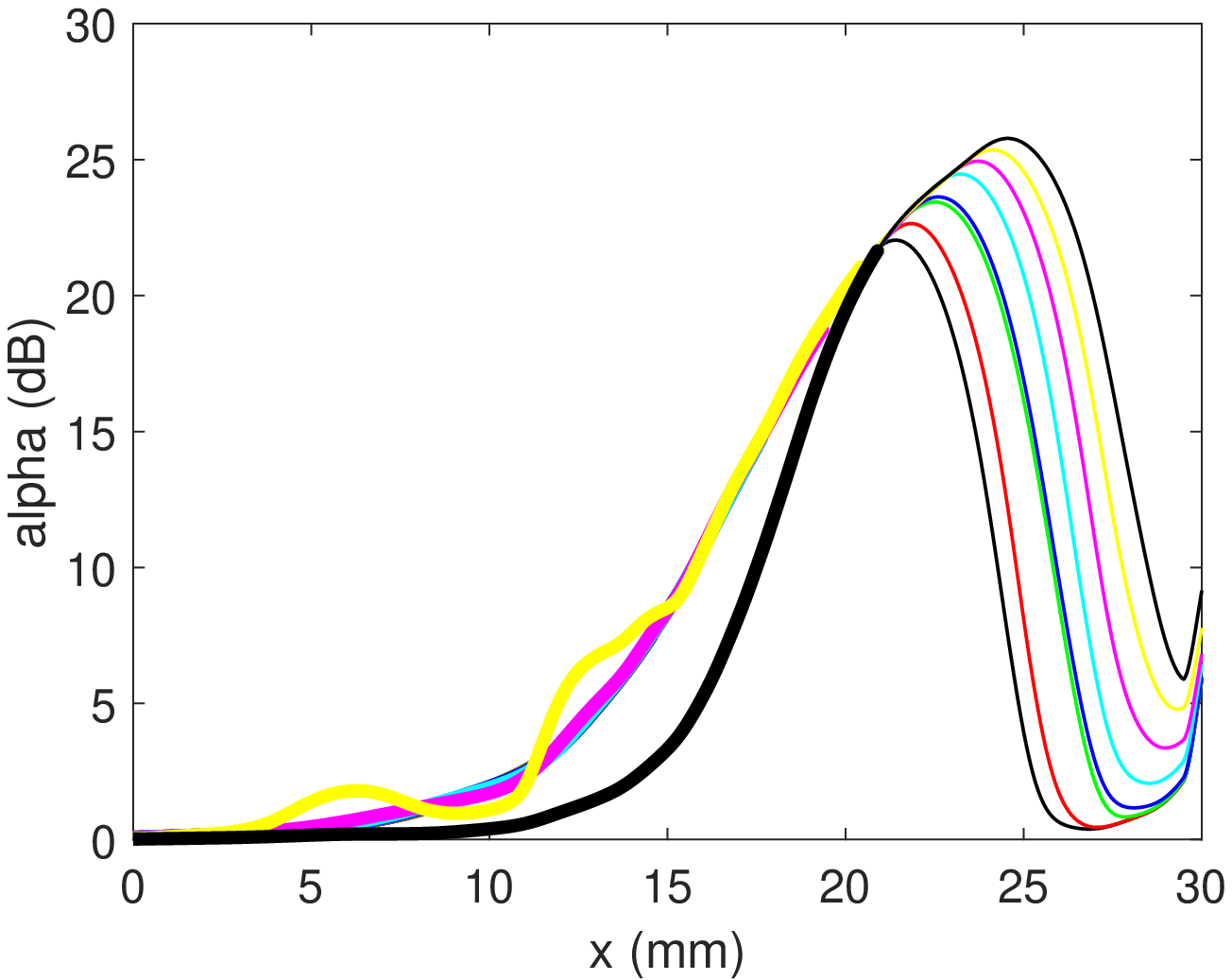}\caption{\label{fig:2e}}
\end{subfigure}
\begin{subfigure}{.32\textwidth}
\includegraphics[width=\linewidth]{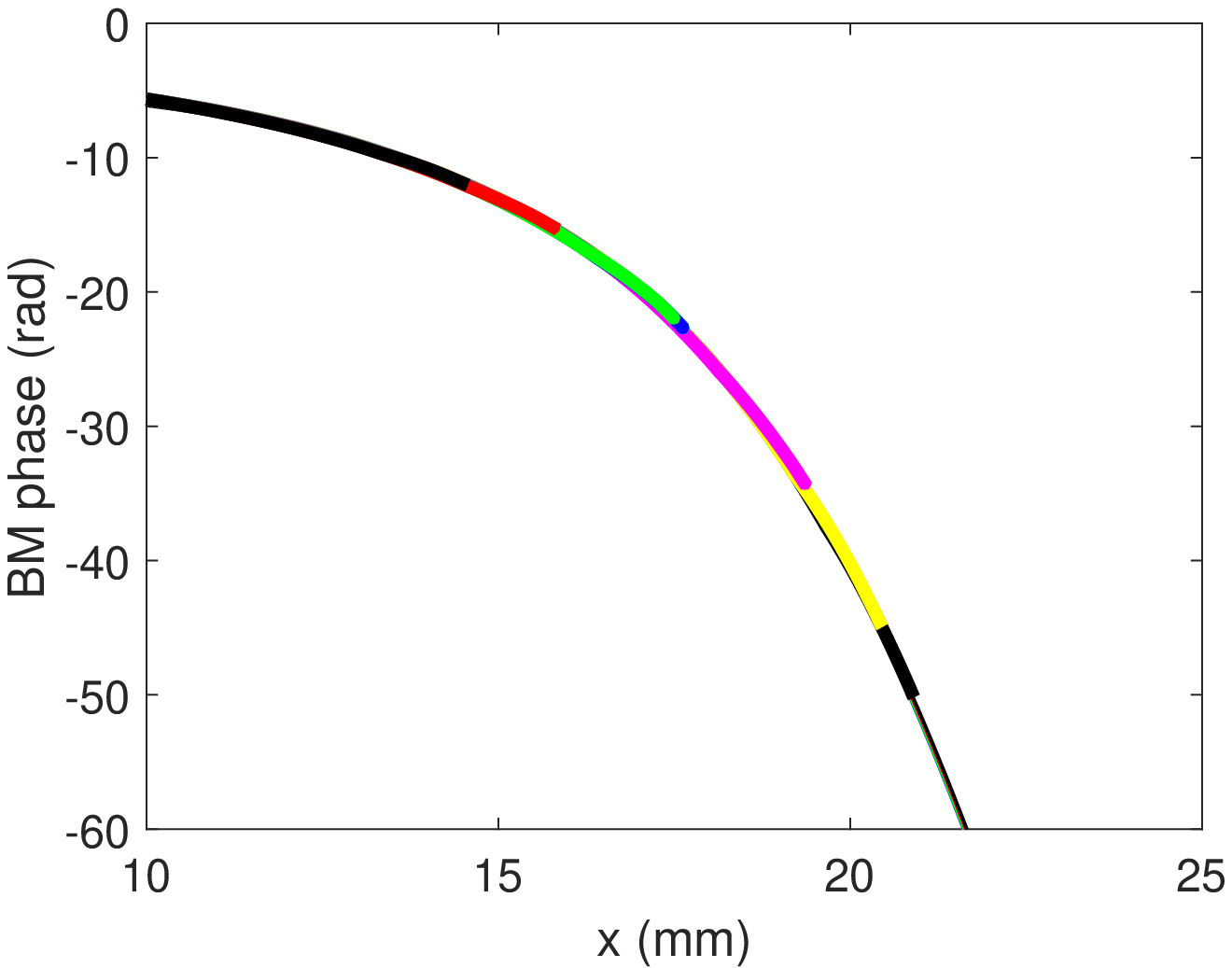}\caption{\label{fig:2f}}
\end{subfigure}
\caption{\label{fig:2}Response of the FE model for a sinusoidal stimulus a frequency of \SI{2200}{\hertz}, with $\mu=10\mu_{\text{water}}$; the active term factor $G$ is varied from \num{1} to \num{4.5} with steps \num{0.5}: (a) BM velocity profile; (b) spatial profile of the differential pressure “measured” near the BM ($z_{\pm}=\SI{\pm15}{\micro\meter}$); (c) local admittance; (d) wavenumber computed from the BM response using Eq.~(\ref{eq:k2Shera}); (e) local amplitude of the factor $\alpha$; (f) BM phase. Thick lines in panels (c)-(f) extend from the base to the position of the BM response peak.}
\end{figure}
In Fig.~\ref{fig:2}, the FE model (FEM) results are shown for $\mu = \num{10}\mu_{\text{water}}$, varying $G$ between \num{1} and \num{4.5}, with steps \num{0.5}, to get a dynamic range similar to that of Fig.~\ref{fig:1}. One can immediately notice that, while the longitudinal shift of the peak and the phase lag decreases to more realistic values, the other features of the response show the same behavior as in Fig.~\ref{fig:1}. Therefore, the main results of this study, regarding the role of focusing and the admittance invariance, are not dependent on a specific choice of the viscosity value, and could be considered as roughly representative also of the behavior of more complex models, in which viscoelastic materials are used to model different elements of the OC, and the viscosity losses due to interstitial fluids are also considered.

\begin{figure}[ht]
\centering
\begin{subfigure}{.32\textwidth}
\includegraphics[width=\linewidth]{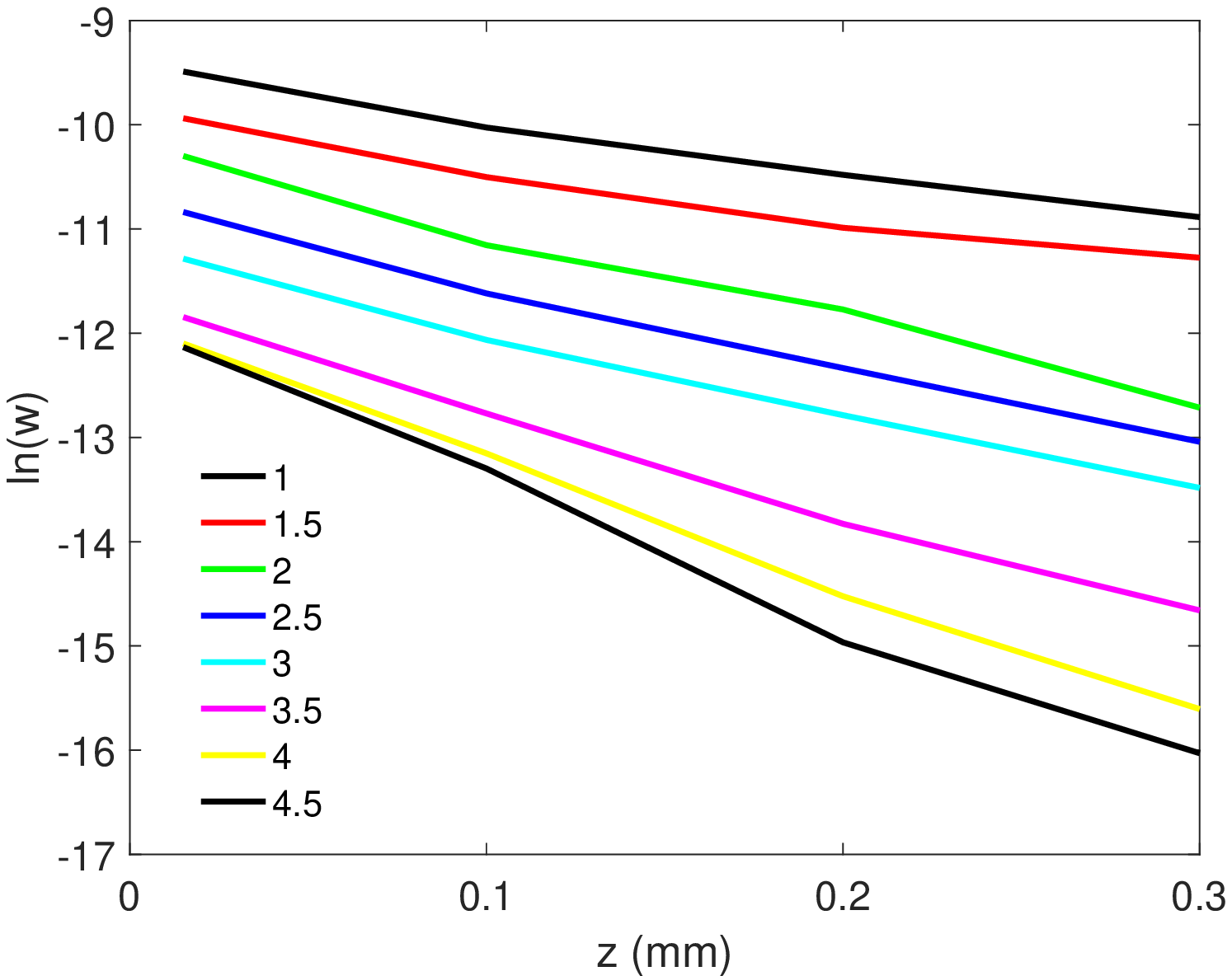}\caption{\label{fig:3a}}
\end{subfigure}
\begin{subfigure}{.32\textwidth}
\includegraphics[width=\linewidth]{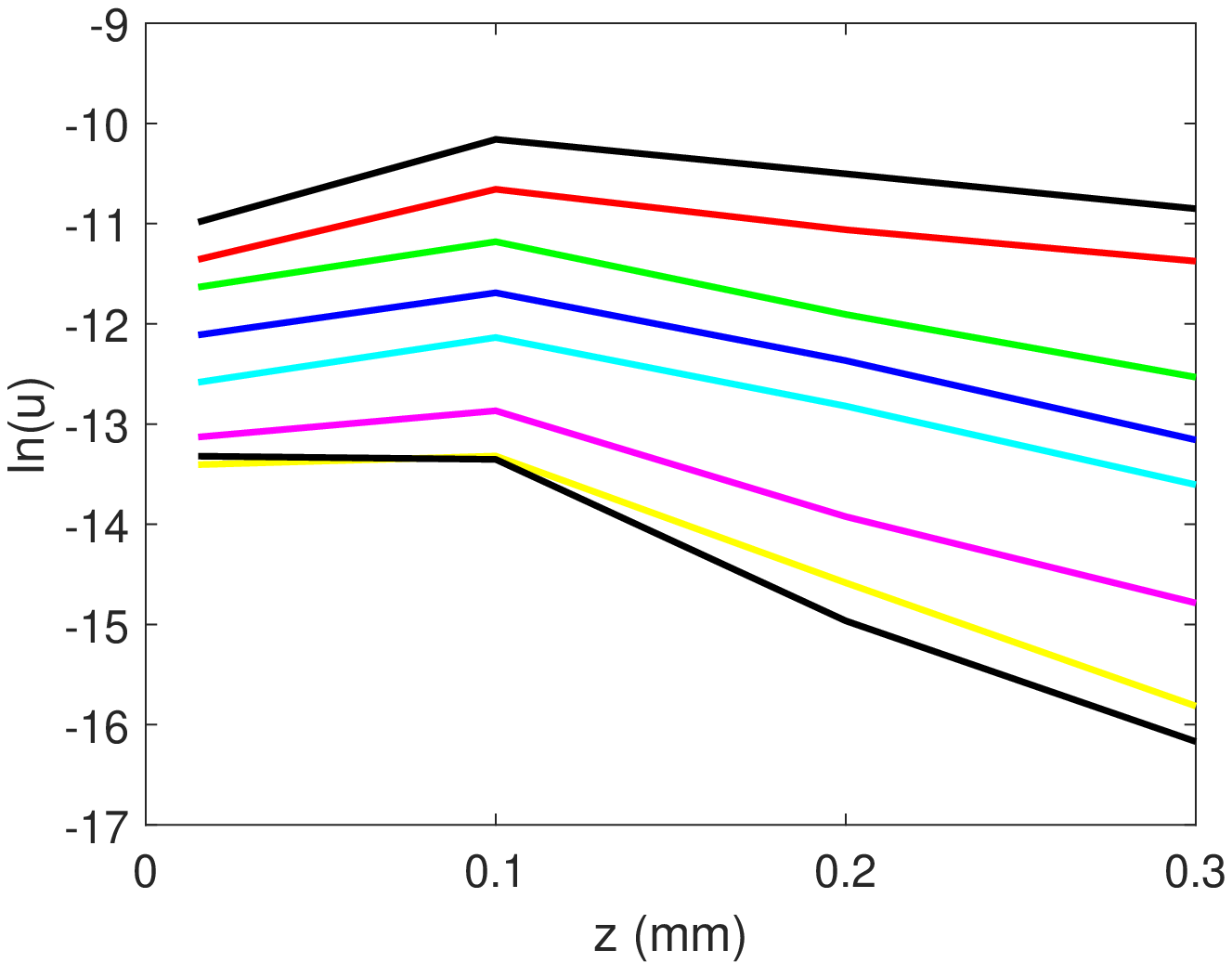}\caption{\label{fig:3b}}
\end{subfigure}
\caption{\label{fig:3}Vertical profiles of the $w$ and $u$ velocity components, directed, respectively, along the $z$- and the $x$-axis, computed for different values of $G$, at the position of the response peak. Log units are used to allow a visual estimate of the slope, which is approximately equal to the local wavenumber.}
\end{figure}
The focusing effect is well visible in Fig.~\ref{fig:3}, where the vertical profiles of the $w$ and $u$ components of the velocity field are plotted, for each value of $G$, at the position of the corresponding response peak.
The logarithm of the velocity is plotted to allow one to estimate the value of the wavenumber from the slope of the curves. Indeed, in the short-wave region, the $\sinh$ and $\cosh$ functions describing the profiles of $w$ and $u$, respectively, are well approximated by functions proportional to $e^{-kz}$. The values of the wave vector at the peak estimated from the slopes are consistent with the estimate obtained from Eq.~(\ref{eq:k2Shera}) and shown in Fig.~\ref{fig:2d}. One may also appreciate that the exponential law is actually well verified by $w$, whereas $u$ starts to follow the same behavior only for $z > \SI{100}{\micro\meter}$, because in a thin layer close to the BM the fluid velocity vector is parallel to $z$ due to the boundary conditions imposed by viscosity.

\begin{figure}[ht]
\centering
\begin{subfigure}{.32\textwidth}
\includegraphics[width=\linewidth]{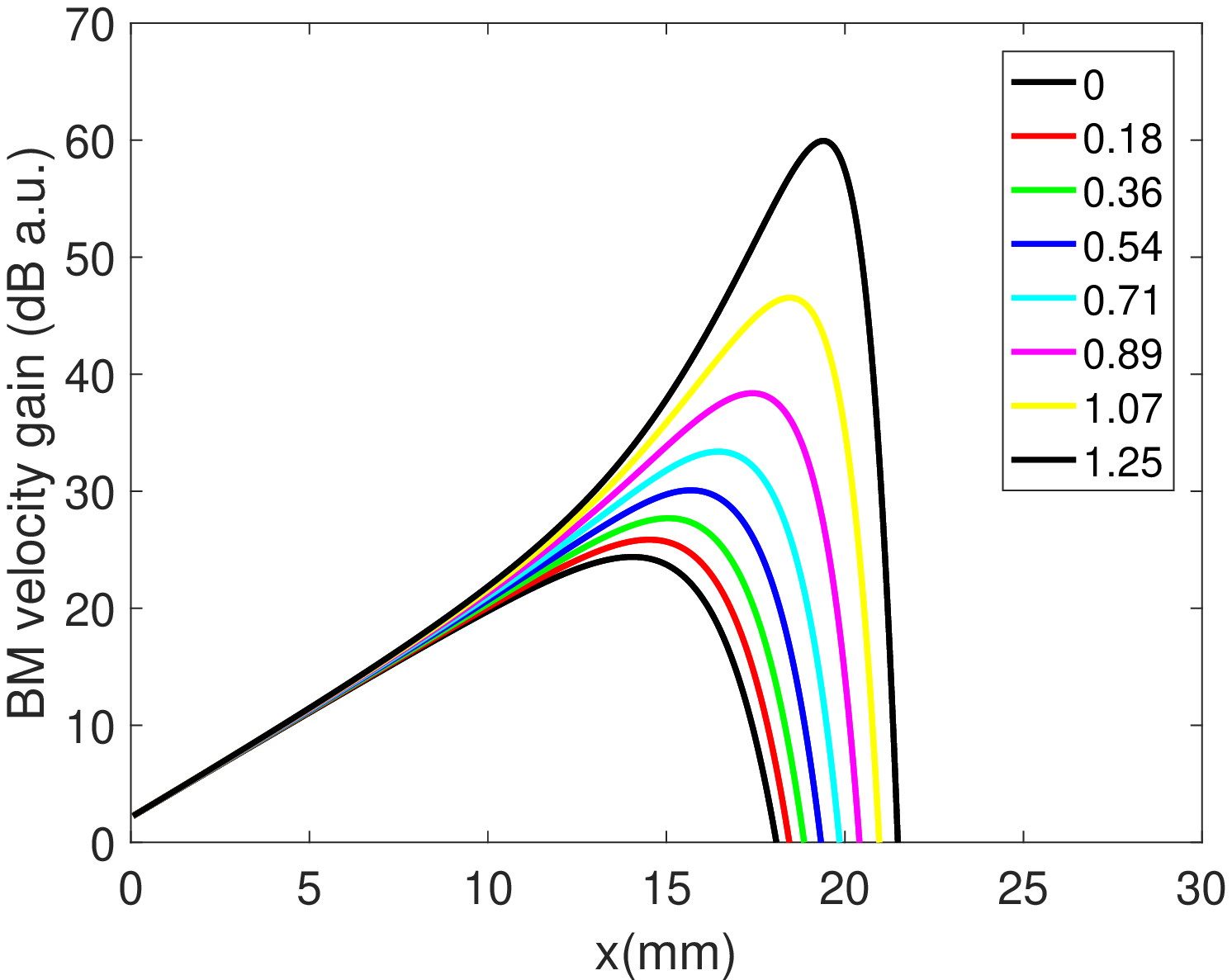}\caption{\label{fig:4a}}
\end{subfigure}
\begin{subfigure}{.32\textwidth}
\includegraphics[width=\linewidth]{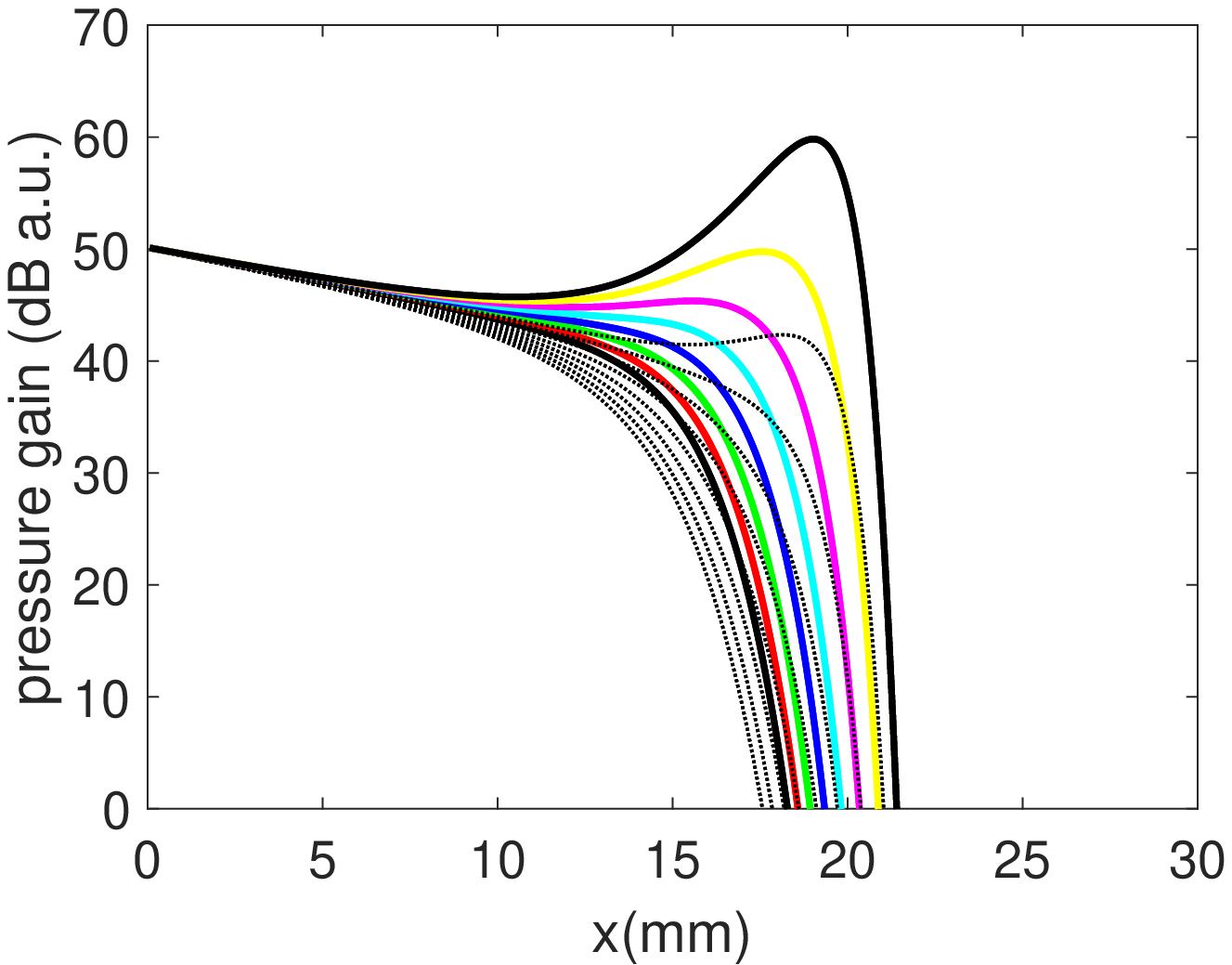}\caption{\label{fig:4b}}
\end{subfigure}
\begin{subfigure}{.32\textwidth}
\includegraphics[width=\linewidth]{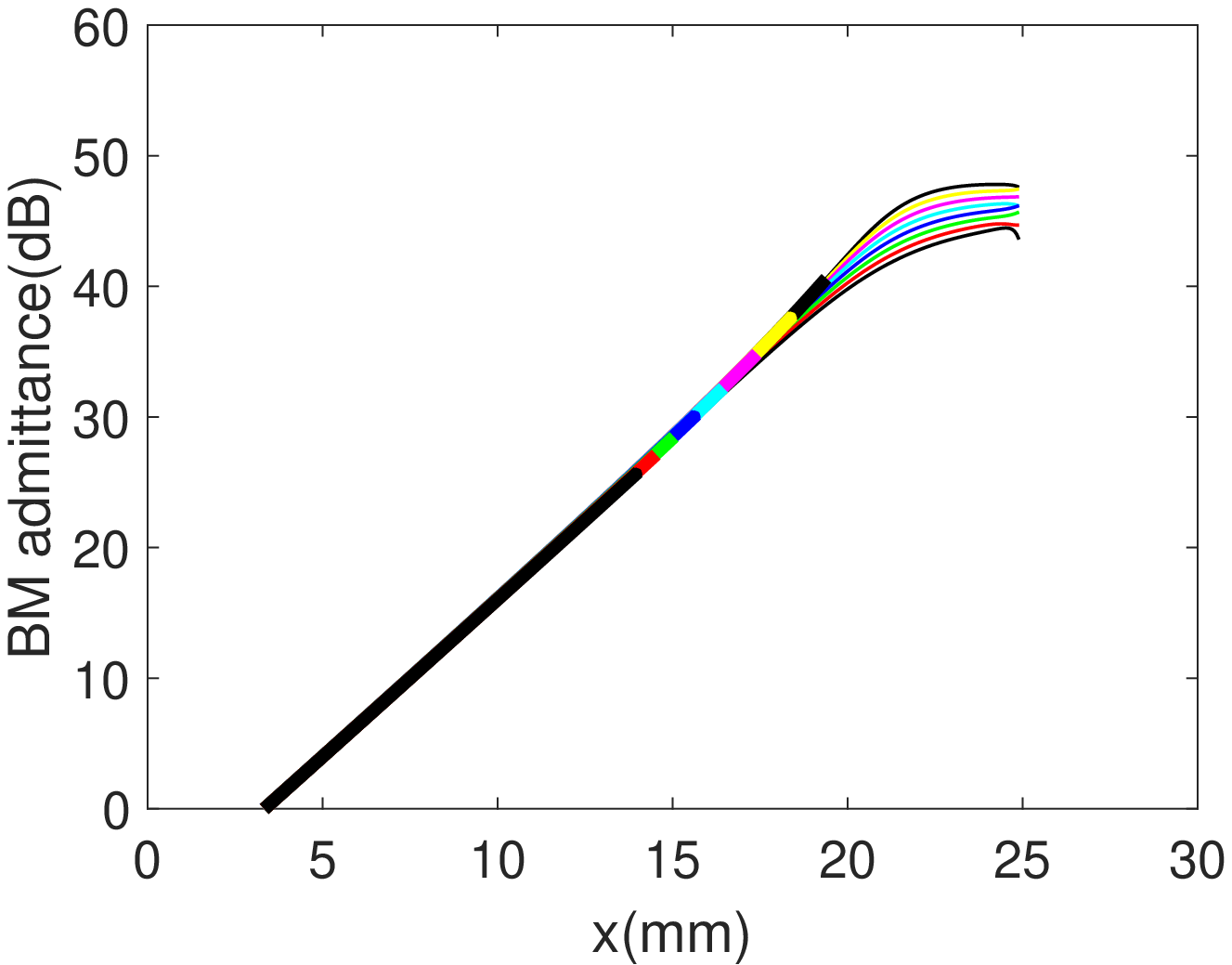}\caption{\label{fig:4c}}
\end{subfigure}
\begin{subfigure}{.32\textwidth}
\includegraphics[width=\linewidth]{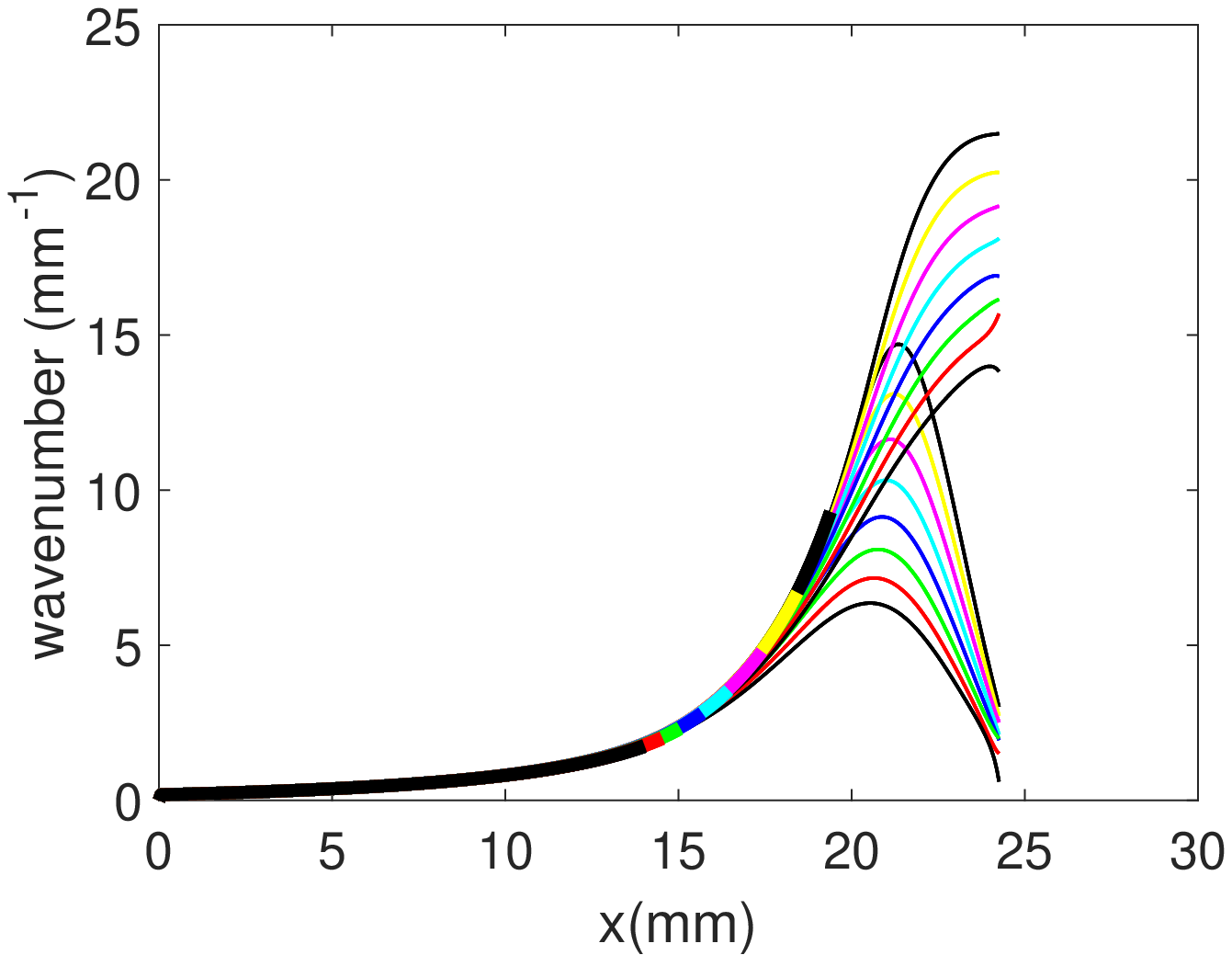}\caption{\label{fig:4d}}
\end{subfigure}
\begin{subfigure}{.32\textwidth}
\includegraphics[width=\linewidth]{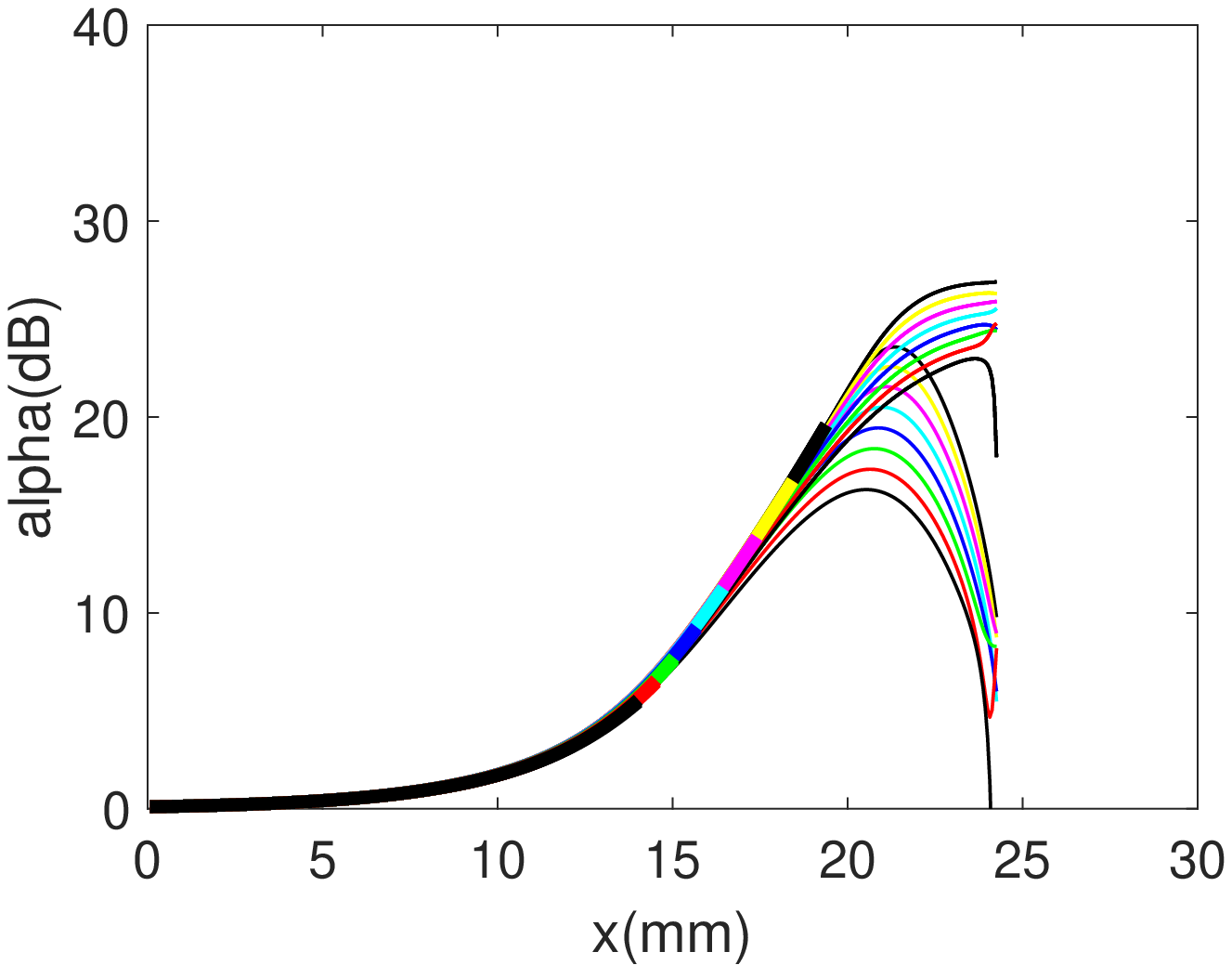}\caption{\label{fig:4e}}
\end{subfigure}
\begin{subfigure}{.32\textwidth}
\includegraphics[width=\linewidth]{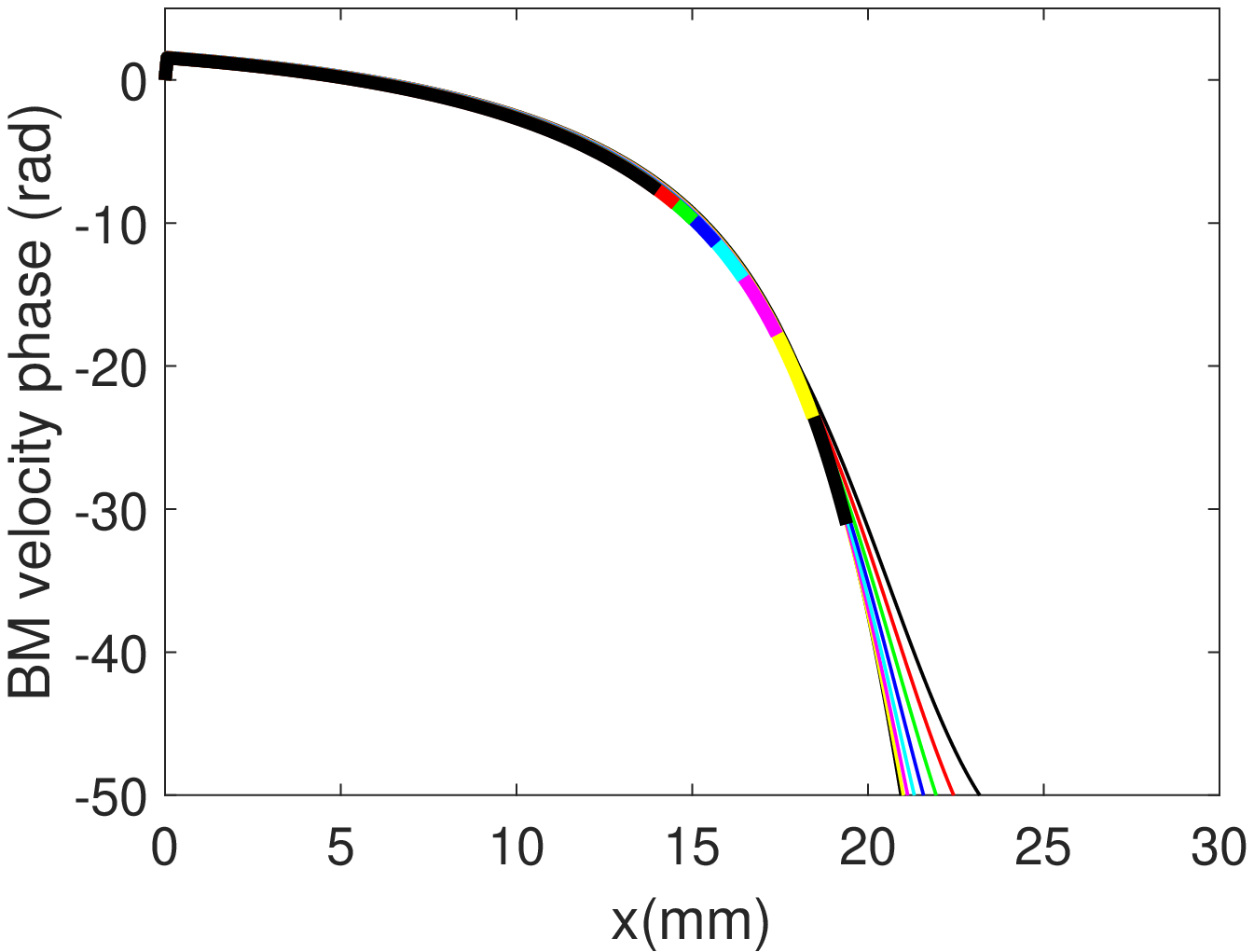}\caption{\label{fig:4f}}
\end{subfigure}
\caption{\label{fig:4}2-D model solutions using the WKB approximation, with $\mu=10\mu_{\text{water}}$; the active term factor $G$ is varied from \num{0} to \num{1.25}: (a) BM velocity response for a viscous fluid model for different values of the anti-damping term; it can be noted that, with increasing anti-damping, the transition between low and high gain is smooth, and the model is never unstable, whereas in an anti-damping model without fluid viscosity the transition between low gain and instability would be unnaturally sharp; (b) fluid pressure gain (in color: near the BM; dotted lines: average pressure); (c) BM admittance; in a viscous fluid model, the increasingly active term injects the power necessary to sustain a large value of the wavenumber, compensating the viscous losses, without a significant increase of the admittance; (d) local amplitude (and real part) of the wavenumber computed from the iterative solution; (e) factor $\alpha$; (f) BM phase profile. Thick lines in panels (c)-(f) extend from the base to the position of the BM response peak.}
\end{figure}
Results obtained from the 2-D model solved using the WKB approximation are shown in Fig.~\ref{fig:4}, for $\mu = \num{10}\mu_{\text{water}}$, varying $G$ in the range \numrange{0}{1.25}. The FEM results are reproduced quite nicely, with a smaller progressive apical shift of the response peak with increasing $G$, whose range was set in order to get the same dynamics of the FEM. The BM velocity and pressure response of the model show a large gain dynamical range in the peak region (Fig.~\ref{fig:4a},\ref{fig:4b}), while the local admittance shows little variation among the models of very different active term (Fig.~\ref{fig:4c}), compared to the variation of both the pressure and BM velocity responses. Note also that this variation is largest in the region beyond the peak, where both responses drop by orders of magnitude. This behavior is accompanied by small variation of the local value of the wavenumber (Fig.~\ref{fig:4d}), and of the factor $\alpha$ (Fig.~\ref{fig:4e}). On the other hand, as in the FEM simulations, the values of admittance, wavenumber and $\alpha$ at the peak of the response (end of thick lines) are sensitive to $G$. The WKB model also shows a strong dependence on $G$ of the BM phase lag and of the phase slope (Fig.~\ref{fig:4f}) in the peak region. If one performs the simulations using the viscosity coefficient of water, the same BM and pressure amplitude responses as in Fig.~\ref{fig:4} (not shown for brevity) are obtained for $G$ variable in the range \numrange{0}{1.05}. The main difference, as in the FEM simulations, is the value (about twice as large as the one for $\mu=10\mu_{\text{water}}$) of the phase lag and slope at the peak and of the apical shift of the peak position.
\begin{figure}[ht]
\centering
\begin{subfigure}{.32\textwidth}
\includegraphics[width=\linewidth]{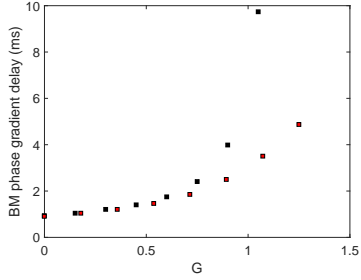}\label{fig:5}
\end{subfigure}
\caption{\label{fig:5}Phase gradient delay at the peak, as a function of $G$, for $\mu=\mu_{\text{water}}$ (black), and $\mu=10\mu_{\text{water}}$ (red).}
\end{figure}
The phase gradient delay at the peak is shown in (Fig.~\ref{fig:5}) for the two different values of viscosity, showing that a small apical shift of the peak response yields unreasonably large phase gradient delays. Such a large variation of the phase gradient delay may be partly due to the linear approximation used in this study. Indeed, models implementing instantaneous nonlinearity typically show reduced nonlinear dependence of the phase gradient delay on the stimulus level \citep{sisto2015decoupling}.

The amplitude of the complex BM response in the WKB approximation is related to four different factors:
\begin{equation*}
|\dot{\xi}(x,\omega)| \propto |Y_{\text{bm}}(x,\omega)| |\alpha(x,\omega)| |f(x,\omega)|e^{\int_0^x\d x'\Im(k(x',\omega))}.
\end{equation*}
The first factor, the local admittance amplitude, depends on the local passive and active contributions to the transverse impedance of the transmission line. However, in the short-wave region, it is proportional to the wavenumber. The factor $|\alpha|$, which is also proportional to the wavenumber, represents the effect of hydrodynamic focusing, the function $|f|$ is a normalization function decreasing with increasing wavenumber, and the path integral of the imaginary part of the wave vector represents the net transmission amplitude gain accumulated by the TW. It is a negative contribution in passive systems, which becomes positive (or less negative) with increasing activity of the OHC mechanism.

\begin{figure}[ht]
\centering
\begin{subfigure}{.32\textwidth}
\includegraphics[width=\linewidth]{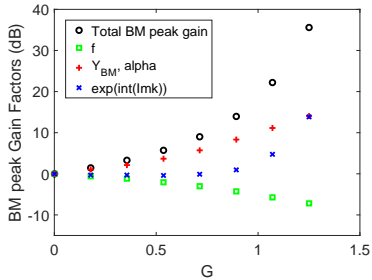}\caption{\label{fig:6}}
\end{subfigure}
\caption{\label{fig:6}WKB Factorization of the total gain dynamics at the response peak as a function of $G$. All the contributions are normalized to their values for $G=0$.}
\end{figure}
The different factors contributing to the BM peak gain are separately plotted in Fig.~\ref{fig:6} as a function of the OHC activity factor $G$, normalized to their respective values computed for $G=0$. Note that the position of the peak moves towards the apex, as in the FEM simulations, so the comparison is not made at the same position $x$. The WKB normalization function $f$ gives an increasingly negative contribution, while the peak admittance grows proportionally to $\alpha$, both approximately proportional to the wavenumber in the short-wave region. The cumulative contribution to the gain associated with the path integral of the imaginary part of the wavenumber gives a significant contribution only for the most active models ($G\geq\num{1}$). The OHC input power is necessary to sustain the response against the viscous losses (increasing with increasing $\Re(k)$), allowing the TW amplitude to keep increasing up to more apical regions, where the wavenumber is larger, thus indirectly boosting both the pressure and the admittance at the TW peak. In less active systems, the TW amplitude starts to decrease at more basal places, where the wavenumber is smaller. This way, a linear mechanism (hydrodynamic focusing), fed by the nonlinear OHC power, yields a strongly nonlinear BM dynamical range, without a correspondingly large variation of the admittance.

It may be interesting to use the FEM results to evaluate the order of magnitude of the power locally injected by the OHCs.
\begin{figure}[ht]
\centering
\begin{subfigure}{.32\textwidth}
\includegraphics[width=\linewidth]{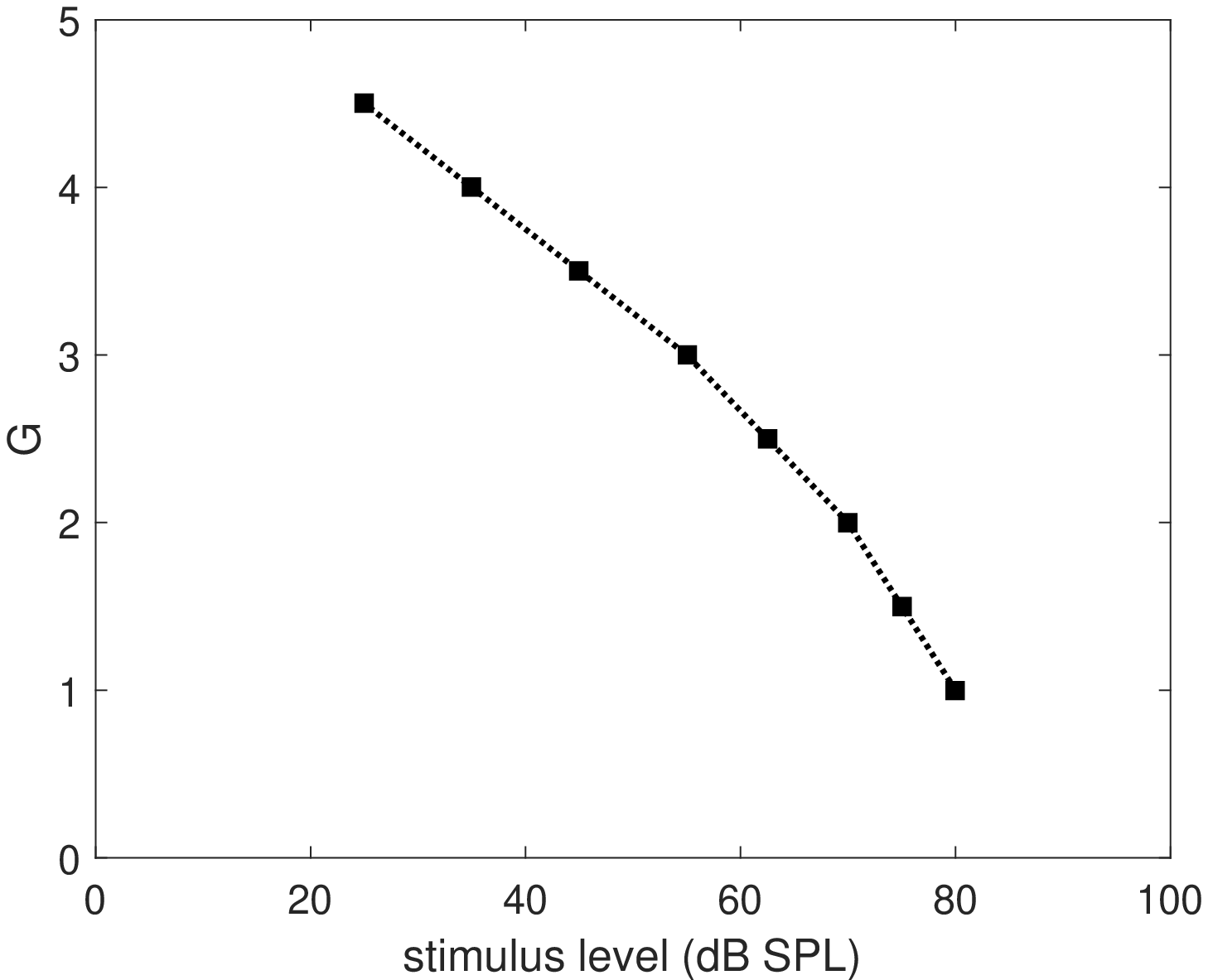}\caption{\label{fig:7a}}
\end{subfigure}
\begin{subfigure}{.32\textwidth}
\includegraphics[width=\linewidth]{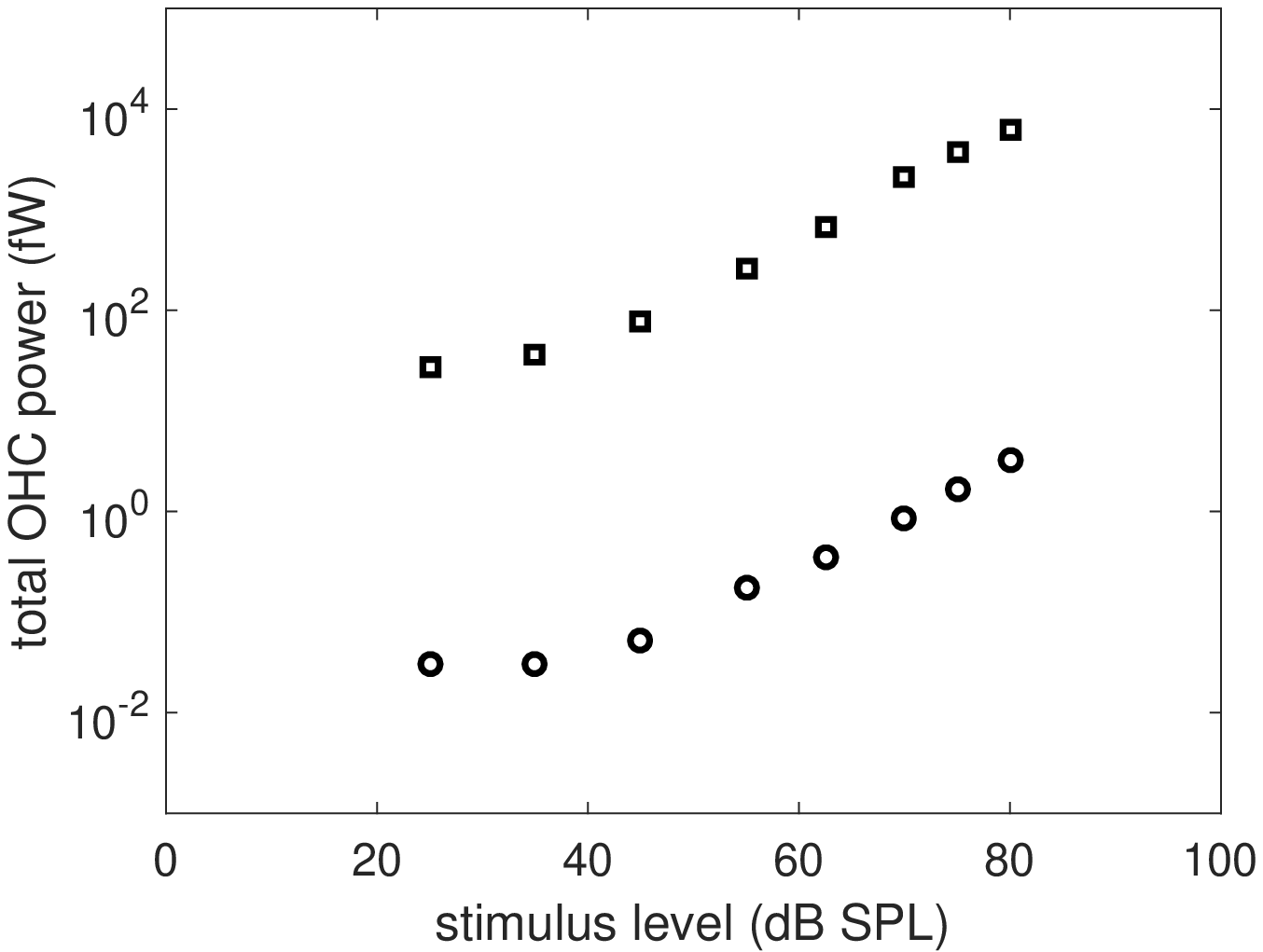}\caption{\label{fig:7b}}
\end{subfigure}
\begin{subfigure}{.32\textwidth}
\includegraphics[width=\linewidth]{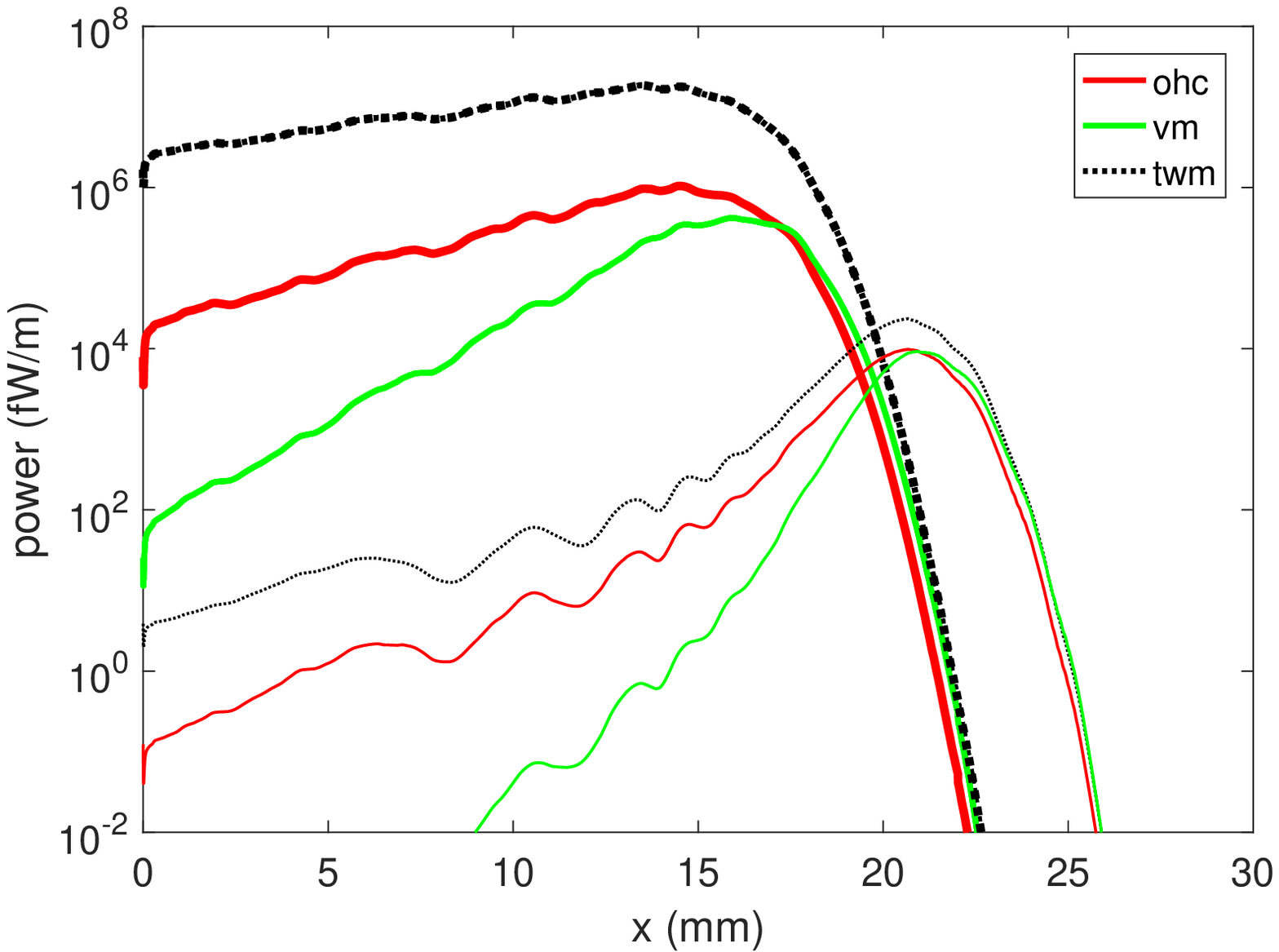}\caption{\label{fig:7c}}
\end{subfigure}
\caption{\label{fig:7}(a) Assumed dependence of the factor G on the stimulus level. (b) Dependence on the stimulus level of the total power input by the OHC mechanism (squares) and of the power from a single OHC (circles). (c) Power per unit length of the TW on the BM (dotted line), of the OHC force (red), of the viscous BM-fluid interface losses (green).}
\end{figure}
As the model is linear, one has to assume a correspondence between the stimulus level and the active factor $G$, as shown in Fig.~\ref{fig:7a}. This dependence is similar to that assumed by \citet{wang2016cochlear}, who performed analogous computations in a 3-D model. We assumed that the active factor $G$ drops more rapidly in the high stimulus level range, and that the highest gain factor used in our simulations (\num{1.25}) corresponds to \SI{25}{\decibel} SPL. With these arbitrary yet reasonable assumptions, and considering \num{11000} OHC distributed over $L=\SI{32}{\milli\meter}$, the computed single OHC power output varies in our simulations between 0.03 and 3 fW, increasing with stimulus level (open circles in Fig.~\ref{fig:7b}), similarly to what was found by \citet{wang2016cochlear}, and compatible with the upper bound they set at about \SI{30}{\femto\watt}. The total OHC power integrated along the BM is shown by the open squares. In Fig.~\ref{fig:7c} we plotted the power per unit length of the TW on the BM (dotted black line), compared to the OHC power (red) and the power dissipated by the viscous force on the BM (green), for $G=0$ (80 dB, thick lines) and $G=1.25$ (25 dB, thin lines). One may note that the dissipated power, being propotional to the wavenumber, grows steeply approaching the peak of the response, whose position corresponds to a condition of approximate equality between power input and output.

\section{Discussion}

Using analytic and numerical methods in 2- and 3-D models of the cochlea, we have shown that the additional spatial dimension is necessary to account for two important hydrodynamic phenomena: the focusing of the pressure wave in a thin layer near the BM and the development of stabilizing viscous forces at the BM-fluid interface.

The high value of the focusing factor $\alpha$ at the peak, and its dependence on the active term effectiveness $G$ demonstrate, respectively, that the focusing phenomenon provides indeed a significant contribution to the overall pressure and BM gain, and to its wide dynamical range, because the variation of $G$ mimics the nonlinear variation of the OHC nonlinear active response.

The onset of both focusing and viscous damping is driven by the development of a sharp BM response (short wavelength), which would not be possible in the absence of an active OHC mechanism increasing the passive peak admittance, thus the OHC force remains the “primum mobile” even in this modified view. Moreover, our simulations also show that, once the focused regime is entered, the maximal active force effectiveness determines the peak amplitude and the longitudinal position of the TW peak, through a competition with the viscous damping forces, which also increase with increasing wavenumber. In the short-wave limit, the focusing mechanism strongly amplifies the pressure near the BM, and, consequently, the BM velocity response, without changing directly the local admittance \citep{shera2005coherent}, but changing the position of the peak, and, therefore, the peak admittance. The fluid viscosity has an important stabilizing effect, because it rapidly drags power from the wave in the region close to the BM as the transverse fluid velocity profile gets sharper due to the synergistic effects of OHC anti-damping and pressure focusing. 

Taken together, focusing and viscosity make the responses of the cochlear model much less sensitive to the fine-tuning of the active force (indeed, in our FEM and WKB simulations the net damping on the BM reaches negative values in the most active cases, which would mean instability, without viscosity, even without focusing), and they yield sharp profiles at low stimulus levels and large gain dynamics for both pressure and BM velocity, with a moderate nonlinear change of the admittance. As already observed by \citet{prodanovic2019power}, viscosity may paradoxically improve the cochlear tuning, because it helps suppressing the response within a narrow spatial region beyond its peak.

The two hydrodynamic effects, focusing and viscous damping, modify the behavior of the TW in the peak region with respect to that associated with a traditional locally-resonant transmission-line active model. In that case, each frequency component of the TW grows approaching its own resonant place, with the OHC forces providing a region of negative damping (with positive imaginary part of the wavenumber) just before the peak. In such a traditional model, the apical cutoff of the TW is at the resonant place, where the imaginary part of the transverse impedance changes sign. In the model of the present study, the TW amplitude is amplified by focusing due to the large value of $\Re(k)$, and then suppressed by fluid viscous damping, due, again, to the large value of $\Re(k)$, well before reaching the resonant place. The OHC forces are necessary to sustain the large value of $\Re(k)$ against the viscous losses. In more active models (lower stimulus levels) they do it more effectively, allowing the “tall and broad” peak of the response to grow higher and get closer to the nominal resonant place, which, in the most active simulations of our FEM and WKB models is at about $x=\SI{25}{\milli\meter}$.

The progressive stabilization due to fluid viscosity is intrinsically different from that due to the saturation of the OHC active force in a system which would be linearly unstable at lower displacement levels. In the latter case, the solution grows in the time domain until it reaches a saturation displacement level, independent of the stimulus level, and then saturates in a rather abrupt way. The fluid viscous damping, being proportional to the wavenumber, counteracts the focusing effect within the same cochlear region, yielding a softer compressive saturation of the BM response.

Although the comparisons of this study between the numerical and analytical WKB simulations suggest that the main physical aspects of this phenomenology have been correctly identified by the 2-D analytical schematization, we must remark that equivalent response was obtained in the FEM and WKB models using quite different ranges for the parameter $G$, associated with the OHC input power. Another limitation of the study, which could alternatively be considered as a useful indication for further studies, is the necessity of using a viscous coefficient much larger than that of water to get reasonably small apical shift of the response peak in differently active models (i.e., as a function of the stimulus level), and reasonably small phase gradient delays. Taken together, these two observations suggest that some quantitative aspects of the dissipative physical mechanisms within the real OC are still outside our oversimplified representations (both FEM and WKB).   

\subsection{Comparison with previous works on fluid viscosity}

If one neglects viscosity, the problem of stability in a system in which anti-damping forces are involved in the gain generation may be considered as a possible flaw of the models schematizing the cochlear gain as due to a large variation of the admittance. The problem of the stability becomes more serious when a 2 or 3 dimensional fluid coupling is considered. In other words, if the realistic cochlear hydrodynamics is kept into account, the fluid focusing makes the instability risk more troublesome.

The role of viscosity has been acknowledged in several studies as a very important one. In the model by \citet{nobili1998well}, in which the 2-D hydrodynamics is accounted for by the Green functions for the pressure, and fluid focusing is effective, a viscous term was proposed:
\begin{equation*}
s_{i+}\left(\frac{\d\xi_{i+1}}{\d t}-\frac{\d\xi_i}{\d t}\right) + s_{i-}\left(\frac{\d\xi_{i-1}}{\d t}-\frac{\d\xi_i}{\d t}\right),
\end{equation*}
representing viscous forces acting on the single oscillator due to the fact that each BM element moves with a different velocity with respect to the adjacent sites. As the in the model by \citeauthor{nobili1998well} the pressure is solved implicitly, the shear velocity was considered as a force acting between the cochlear partitions. The viscous force in \citeauthor{nobili1998well} is proportional to $k^2$ instead of $k$, but its effect on stability is quite similar. In our model we preferred to consider the cochlear partitions as decoupled longitudinally, considering the damping force as due to the velocity gradient at the fluid-membrane interface.

\citet{steele1999cochlear} have considered the fluid viscosity in the full 3-D model of the cochlea, and the effect of the fluid viscosity was further analyzed in \citet{wang2016cochlear}. Using a 3-D box model of the cochlea including the fluid viscosity on the BM and in the bulk, the authors compared the viscous losses to the power generated by the OHCs, with results consistent with those of the present study. In that paper the viscosity is necessary for the power balance, but the problem of the stabilizing force is not addressed. We used the \citeauthor{steele1999cochlear} expressions to compute the correction to the factor $\alpha$ associated with the dissipative Laplacian term in the Navier-Stokes equations, obtaining:
\begin{equation*}
\alpha_{\text{corr}} = \frac{kH}{\tanh(kH)}\left(1-k\sqrt{\frac{\mu}{\i\omega\rho}}\right),
\end{equation*}
which implies reduced effectiveness of the focusing mechanism in the apical cochlea.

The implementation of the viscosity into the scalae lymph in a FEM model was performed by \citet{soons2015basilar}. In that model, developed in Comsol Multiphysics, the viscosity effect was kept into account only in a region very close to the BM. In this region (\SI{\sim20}{\micro\meter}), the linearized Navier-Stokes equations are solved whilst the contribution of the viscosity is neglected inside the fluid volume to reduce the model computational cost. The role of the viscosity seems to be relevant, especially, again, to the aim of explaining the power balance inside the cochlea. In \citet{sasmal2019unified} a model including a realistic FE representation of the OC is proposed, in which the viscosity is also present, and they acknowledge the stabilizing role of viscosity as a key ingredient of any cochlear model. Their model makes use of a Newton equation at the interface including a viscous force on the fluid boundary layer that is qualitatively different, but formally very similar (Eq.~S2 of their SI Appendix) to that of the present paper. Their conclusion that viscosity is more effective in the apex due to the larger thickness of the boundary layer seems to be contradicted by the explicit solution proposed by \citet{steele1999cochlear}, which shows that the variation of the thickness of the layer is compensated by that of the amplitude of the force.

\section{Conclusions}

Two important effects of the fluid hydrodynamics, pressure focusing and viscous dissipation, account for the experimentally observed high gain and stability of the BM response, as demonstrated by numerical simulations of a 3-D FE cochlear model, and confirmed and explained by a WKB 2-D model, in which analytical expressions for the two effects are used.

Several aspects of the computed BM and differential pressure response are discussed, highlighting the agreement between the results of the two formulations and several features of the experimental BM and pressure response, which supports the approximate validity of the analytical assumptions of the 2-D WKB formulation.

\section*{Acknowledgement}
We thank Christopher A. Shera and Alessandro Altoè for helpful suggestions and collaboration in the early stages of this study.

This work was supported by INAIL grant BRiC 2019 ID9/2019.

\bibliographystyle{abbrvnat}
\bibliography{Fluidpaper}

\end{document}